\documentclass[12pt]{article}

\usepackage{amsmath,amsfonts,amssymb,color,graphicx,overpic,}
\usepackage{slashed}
\usepackage{epsfig}
\usepackage[utf8]{inputenc}
\usepackage{amsmath}
\usepackage{ wasysym }
\usepackage{graphicx}
\usepackage[english]{babel}
\usepackage{graphic x}
\usepackage{relsize}
\usepackage{cite}

\begin{document}

\begin{titlepage}
\rightline{Jan 2018}
\vskip 1.9cm
\centerline{\large \bf
Dissipative dark matter halos: The steady state solution II}

\vskip 2.1cm
\centerline{R. Foot\footnote{
E-mail address: rfoot@unimelb.edu.au}}

\vskip 0.4cm
\centerline{\it ARC Centre of Excellence for Particle Physics at the Terascale,}
\centerline{\it School of Physics, The University of Sydney, NSW 2006, Australia}
\vskip 0.1cm
\centerline{and}
\vskip 0.1cm
\centerline{\it ARC Centre of Excellence for Particle Physics at the Terascale,}
\centerline{\it School of Physics, The University of Melbourne, VIC 3010, Australia
}

\vskip 2.0cm
\noindent
Within the mirror dark matter model and dissipative dark matter models in general, 
halos around 
galaxies with active star formation (including spirals and gas rich dwarfs)
are dynamical: they
expand and contract in response to 
heating and cooling processes. 
Ordinary Type II supernovae (SN) can provide the dominant heat source,
possible if kinetic mixing interaction exists with strength $\epsilon \sim 10^{-9} - 10^{-10}$.
Dissipative dark matter halos can be modelled as a fluid governed by Euler's equations.
Around sufficiently isolated and unperturbed galaxies 
the halo can relax to a steady state configuration, where 
heating and cooling rates locally balance and hydrostatic equilibrium prevails.
These steady state conditions can be solved to derive 
the physical properties, including the halo density and temperature
profiles, for model galaxies.
Here, we have considered idealized spherically symmetric galaxies within 
the mirror dark particle model, as in the earlier paper 
[Paper I, arXiv:1707.02528],
but we have assumed that the local halo heating in the SN vicinity dominates over radiative sources.
With this assumption, physically interesting steady state solutions arise which we compute
for a representative range of model galaxies. 
The end result is a rather simple description of the dark matter 
halo around idealized spherically symmetric systems, characterized in principle
by only one parameter,
with physical properties
that closely resemble the empirical properties of disk galaxies.


\end{titlepage}

\section{Introduction}

There is abundant evidence from galaxy rotation curves that baryons strongly
influence the dark matter distribution around galaxies.
The observed cored profile of gas rich dwarf and low surface brightness galaxies, 
e.g. \cite{154,blok0,blok1,blok2,blok3,littlethings}, 
and baryonic - dark matter
scaling relations, e.g. \cite{tf,btf,btfnew,salucci,DS,Lelli,MC,Lelli2,stacy}, provide reasonably 
direct indications. These observations pose a challenge to the standard picture, 
which assumes that dark matter has only gravitational interactions.

Dark matter might well have nontrivial self interactions, and it is also possible, 
within the parameter space of potential dark matter models, 
that baryons can strongly influence dark matter halo properties via nongravitational 
interactions.  One such framework arises if dark matter is dissipative, as could 
happen if dark matter originates from a hidden sector that contains a 
`dark proton' and a `dark electron', coupled together via a massless 
`dark photon' \cite{sunny1}.  Such dark matter is dissipative in that it can 
cool via the emission of dark photons.  A theoretically constrained particle 
model of this kind arises if the hidden sector is exactly isomorphic to 
the ordinary sector \cite{flv}, a scenario called {\it mirror dark matter} 
(see e.g.\cite{review} for a review and detailed bibliography). 
An important feature of dissipative dark matter is that the
heating is sourced from ordinary supernovae \cite{sph} - 
which thereby supplies the connection with baryons.
This mechanism requires only a tiny coupling between the ordinary and 
dark sector particles, which can be facilitated  by the kinetic mixing interaction \cite{holdom,he}.

In this picture, dark matter halos around 
galaxies with active star formation, including spirals and gas rich dwarfs, 
form a dark plasma. This dark plasma
can expand and contract in response to heating and cooling processes.
Such dark matter can be modelled as a fluid, obeying Euler's equations.
For a sufficiently isolated and
unperturbed galaxy, the dark matter halo can  evolve and approach the
steady state configuration.
In this limit, and assuming that there is no bulk halo motion, 
Euler's equations reduce to two simple conditions: 
the balancing of local heating (${\cal H}$) and cooling rates (${\cal C}$), and 
the pressure gradient with the gravitational force (hydrostatic equilibrium):
\begin{eqnarray}
{\cal H} &=& {\cal C}
\ , \nonumber \\
\bigtriangledown  P &=& - \rho \nabla \phi  
\ .
\label{SSX}
\end{eqnarray}
These equations determine the physical properties of the dark halo, including its 
density, $\rho$.
This is a rather nontrivial and perhaps subtle picture. The fluid dynamics leads to halo evolution
approaching the steady state configuration
in which the halo density profile is strongly influenced by the distribution of supernova (SN) sources -
the primary halo heat source.

This is the general idea. 
There are a number of important and uncertain details yet to be fully understood.
One of these, concerns the mechanism by which ordinary SN transfer energy to the dark halo.
In the SN core, 
the kinetic mixing interaction generates an expanding  energetic plasma 
of dark matter particles and dark radiation,
which interacts with itself and the surrounding halo dark matter. 
Ultimately, this results in two kinds of halo heating: 
A  dark radiation component, transporting energy far from the SN,
and a local heating in the SN vicinity via particle collisions, shock heating etc.
The previous detailed study \cite{footpaperI} (hereafter Paper I), 
only considered heating via dark radiation.
The dark photons are absorbed in the halo via the photoionization of a halo mirror metal component.
It was found, however, that for mirror dark matter, 
such heating turns out to be quite ineffective as
the halo becomes highly ionized, and the dark photons are unable to be efficiently absorbed in the halo.
In fact, no physically interesting halo solutions emerged with energy transported via dark radiation.

In this paper, we solve for steady state solutions within the mirror dark matter particle model, 
assuming idealized spherically symmetric systems as in Paper I, but we include local halo heating processes.
If the heating is sufficiently localized, the 
detailed description of the complex processes involved are not 
required, only the total energy transmitted to the halo per (average) SN, $L_{\rm SN}^{e'}$.
It turns out that this leads to physically interesting steady state solutions 
provided that the local processes dominate halo heating.
It also simplifies the emerging picture, for example, no halo mirror metal content
is required, and there is negligible dependence on SN parameters other than $L_{\rm SN}^{e'}$.
The end result is a rather simple description of the dark matter 
halo around idealized spherically symmetric systems, characterized in principle
by only 1 parameter 
(in addition to the baryonic parameters), 
with physical properties that closely resemble the empirical properties of disk galaxies.

\section{Mirror dark matter}

Dark matter might result from a hidden sector, which is almost decoupled from the 
standard model (SM) of particle physics. 
The most theoretically constrained model of this sort arises when the hidden sector is exactly isomorphic
to the SM \cite{flv}.
In terms of a fundamental Lagrangian, the standard model is extended with an exact copy:
\begin{eqnarray}
{\cal L} = {\cal L}_{\rm SM}(e,u,d,\gamma,...) + {\cal L}_{\rm SM} (e', u', d', \gamma',...)
+ {\cal L}_{\text{mix}}
\ .
\end{eqnarray}
By construction,
no new fundamental parameters have been introduced. The elementary `mirror  particles'
have the same masses as their corresponding ordinary matter counterparts,  and their gauge self interactions
have the same coupling strength as the ordinary matter gauge self interactions. 
Since the mirror particles are described by the same Lagrangian as the standard model,
there will be an entire set of `mirror chemical elements': ${\rm H', \ He', \ Li', \ Be', \ B', \ C',...}$
etc.,  the  properties of which will, of course, be completely analogous to the corresponding ordinary
chemical elements: ${\rm H, \ He, \ Li, \ Be, \ B, \ C,...}$.
A theoretical motivation for extending the standard model in this particular way is that
it allows improper Lorentz symmetries to be respected \cite{flv}.

In addition to gravity, the ordinary and mirror particles are assumed to interact with each
other via the kinetic mixing interaction \cite{holdom,he}:
\begin{eqnarray}
{\cal L}_{\text{mix}} = \frac{\epsilon'}{2} \ F^{\mu \nu} F'_{\mu \nu}
\ .
\label{kine}
\end{eqnarray}
Here, $F^{\mu \nu}$ is the standard $U(1)_Y$ field strength tensor.
The effect of kinetic mixing is to embellish the mirror sector  
particles with a tiny ordinary electric charge: $Q = -\epsilon e$ for the mirror electron, $e'$,
and $Q = \epsilon e$ for mirror proton, $p'$. 
(Note that the parameter $\epsilon$, which is proportional to $\epsilon'$, 
is conveniently taken as the fundamental parameter.)

The mirror baryonic particles can be identified with the  nonbaryonic dark matter 
inferred to exist in the Universe, e.g. \cite{blin,hodges,foot04}. 
It is known that
mirror dark matter closely resembles collisionless 
cold dark matter on large scales, e.g. successfully
reproducing the cosmic microwave background (CMB) anisotropy spectrum \cite{ber1,IV,ber2,footcmb}.
On smaller scales, though, the effects of the self 
interactions, and interactions with baryons facilitated by the kinetic mixing
interaction, lead to very different phenomenology.
The steady state solutions discussed here, representing the current
structural properties of halos around galaxies with active star formation, is a prime
example. A brief discussion of other examples can be found in
Paper I and
for a more comprehensive discussion, \cite{review}.

Although mirror dark matter appears rather special
for a number of reasons, one should, of course, keep in mind
that there are other particle candidates if dark matter is indeed dissipative.
Among these are more generic dissipative models, including
the two component model with dark electron and dark proton discussed in Paper I and
e.g. \cite{sunny1}. 
Other possibilities include 
generalized mirror dark matter models with $N > 1$ additional
isomorphic dark sectors, e.g.\cite{foot86}.
For a significant range of parameters,
each of these particle models would be expected to have 
rotation curve phenomenology broadly similar to the mirror dark matter
case studied here.
\footnote{
An implicit assumption is
that all of the nonbaryonic dark matter in the Universe arises 
from a dissipative dark matter model.
Hybrid dark matter models, where a subdominant dissipative component is 
assumed to exist alongside a dominant collisionless particle component, 
have also been discussed in the literature, e.g. \cite{ddd,dddxyz,ddd2}.
}

\section{Dissipative halo dynamics}

\subsection{Some preliminaries}

In the dissipative mirror dark matter model, the halo around rotationally supported galaxies 
takes the form of a dark plasma with long range interactions resulting in collective behaviour. 
This type of dark matter can be modelled as a fluid,
governed by Euler's equations, and if significant dark magnetic fields are present, by magnetohydrodynamical equations.
Dissipation plays an important role:
in the absence of significant heating the
dark matter halo would collapse to a disk on a timescale shorter 
than the current Hubble time.
However, if a significant heat source exists, such galactic dark matter can form an extended halo.
It happens that Type II supernovae (SN), can provide
a heat source for the dark sector. A substantial fraction of the core-collapse energy 
can be transferred into the production of energetic mirror particles via processes
facilitated by the kinetic mixing interaction.

In this picture dark matter galaxy halos are dynamical, influenced by the dissipative cooling as well
as supernova sourced heating. 
For a sufficiently isolated and unperturbed galaxy, the dark halo
is expected to have evolved, approaching the steady state configuration, 
in which the fluid is in hydrostatic equilibrium, and where heating and cooling rates locally balance.
If this holds true, then by
solving for the steady state configuration, the
current structural properties of 
dark matter halos, including their density and temperature profiles, 
can be determined by the galaxy's baryonic properties (essentially 
the SN abundance and distribution). 

As in Paper I, we make a number of simplifying assumptions.
Principal among these are that
all particle species are
in local thermodynamic equilibrium at a common (spatially-dependent) temperature $T({\bf r})$,
dark magnetic fields can be neglected,
and we consider a strict spherically symmetric system.
The last assumption means that we are studying idealized spherical galaxies, 
where the baryons are spherically distributed rather than the observed disk-like
shape.
Specifically, we consider spherical galaxies with both stellar and gas baryonic components,
with stellar mass density
\begin{eqnarray}
\rho_{\rm baryon}^{\rm stars} (r) = m_{*} \ \frac{e^{-r/r_D}}{4\pi r_D^2 r}  
\ .
\label{doc}
\end{eqnarray}
Here, $m_*$ is the stellar mass parameterized in terms of a stellar 
mass fraction: $m_* = f_s m_{\rm baryon}$.
This baryon density profile arises if one assumes that 
the mass enclosed within any given radial distance, $r$,
is the same as that of a Freeman disk \cite{freeman} within the same radius. 
This condition requires the baryon scale length parameter, $r_D$, in Eq.(\ref{doc}) to
be identical to the disk scale length.
In addition to stars, there is also a baryonic gas component. Generally, the baryonic gas distribution
is observed to be spatially more extended than the stellar distribution.
We model
the gas density, $\rho_{\rm baryon}^{\rm gas}(r)$,  with an exponential profile
of the form Eq.(\ref{doc}), but with
$r_D^{\rm gas} = 3r_D$ and total mass $m_{\rm gas} = (1-f_s)m_{\rm baryon}$ cf. \cite{sal17}.

To find the steady state solution we first need to model the heating and cooling rates.
The cooling processes, bremsstrahlung, line emission and capture processes, 
will be computed as in Paper I.
Of course, these processes depend on the detailed composition of the halo, i.e. the relative abundance
of the various mirror ions, including all their ionization states (computed as in Paper I).
We have assumed  
a mirror helium dominated halo, with $n_{\rm He'}/n_{\rm H'} = 10^{0.68}$, consistent with estimates
from mirror Big Bang Nucleoysnthesis (BBN) \cite{paolo2}.
The halo can in principle contain a metal component, which arises from mirror star 
formation and evolution occurring at an early epoch.
Naturally, the abundance and composition of such a halo metal component 
is highly uncertain, but if significant, may influence our results.
The main effect of a metal component is to modify the cooling rate.

\begin{figure}[t]
  \begin{minipage}[b]{0.5\linewidth}
    \centering
    \includegraphics[width=0.7\linewidth,angle=270]{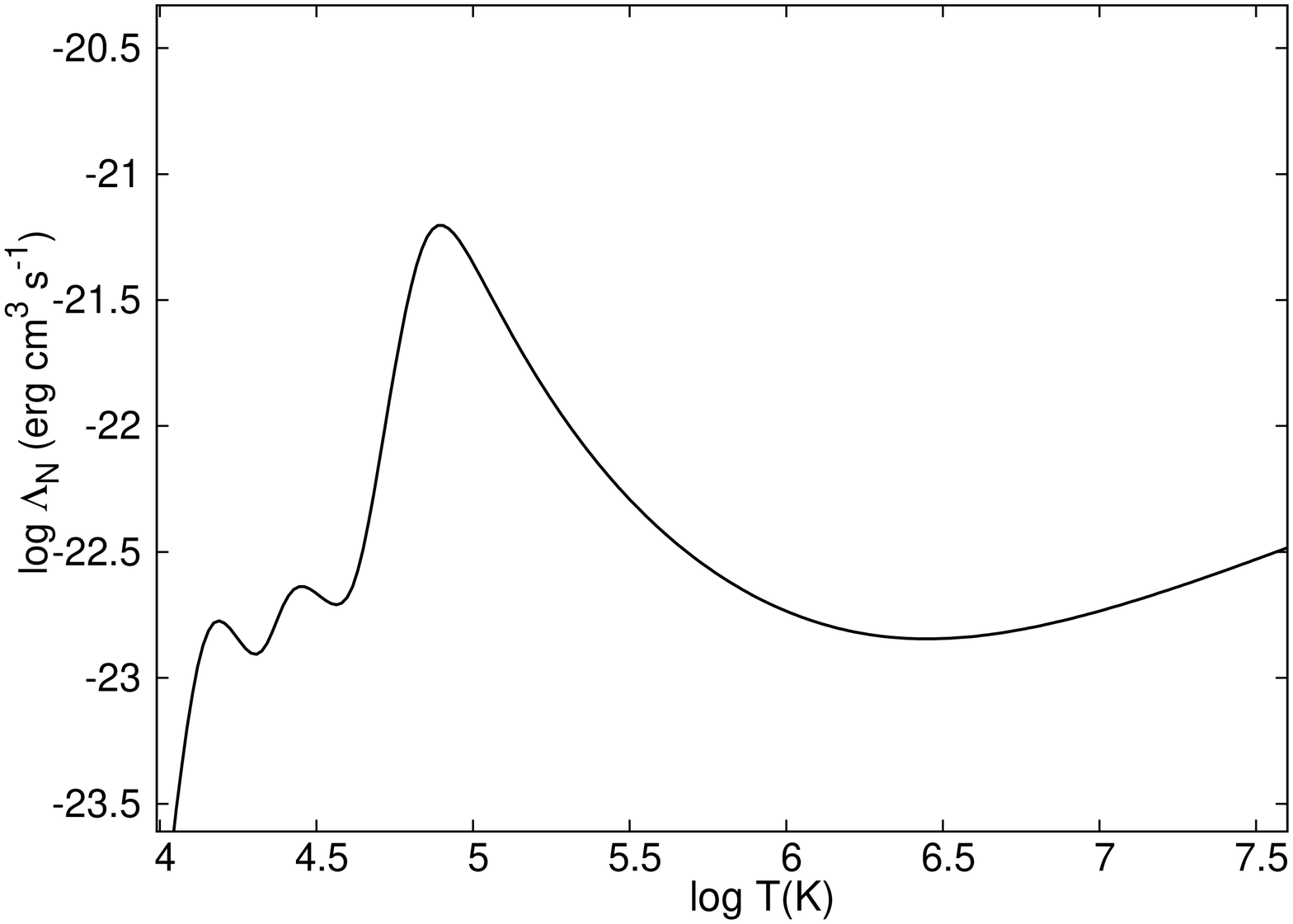}
     (a)
    \vspace{4ex}
  \end{minipage}
  \begin{minipage}[b]{0.5\linewidth}
    \centering
    \includegraphics[width=0.7\linewidth,angle=270]{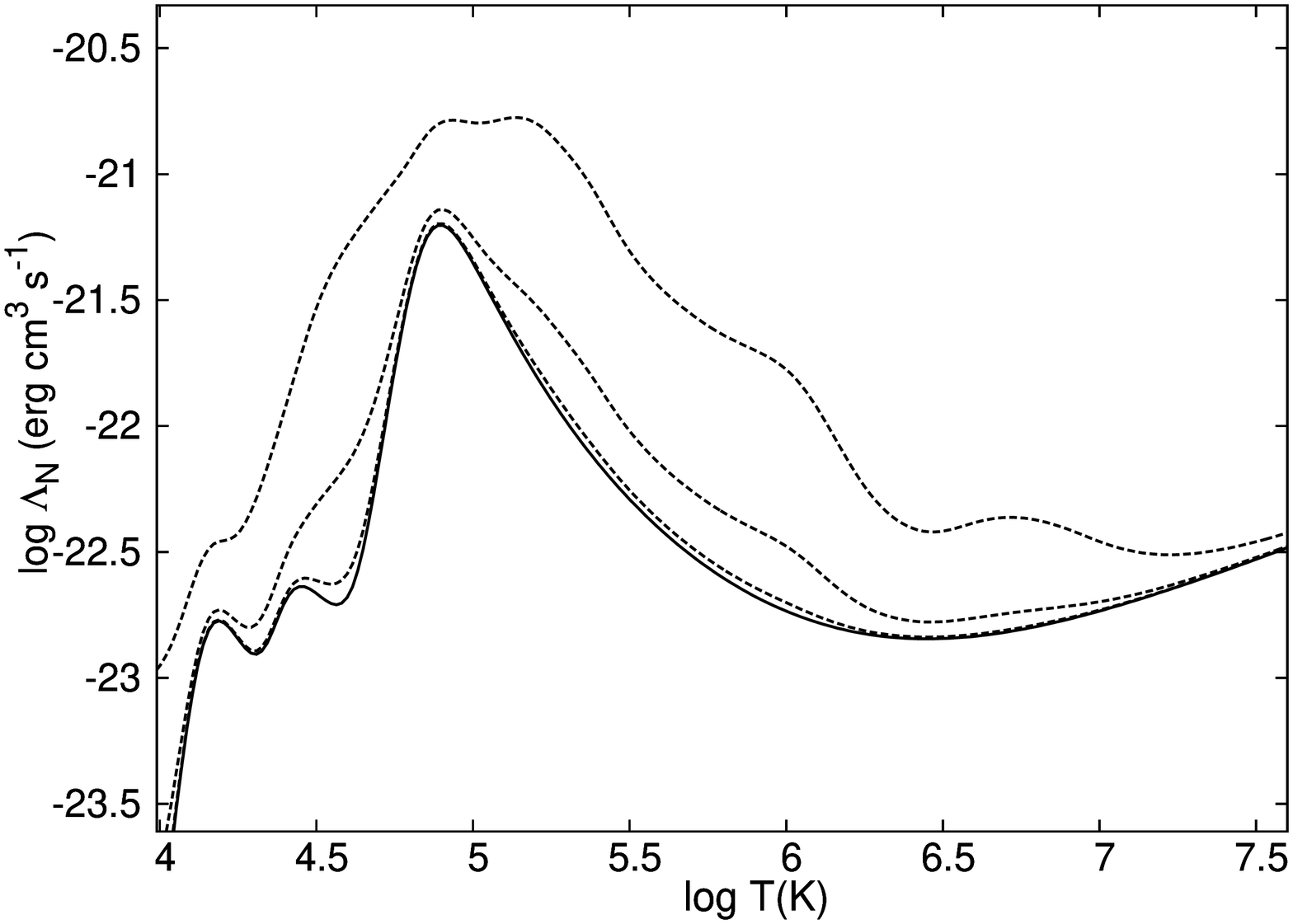}
    (b)
    \vspace{4ex}
  \end{minipage}
\vskip -1.0cm
\caption{\small 
Optically thin cooling function for a mirror helium dominated halo with $\log [n_{\rm He'}/n_{\rm H'}] = 0.68$.
(a) gives the result for a halo consisting of mirror hydrogen and helium (no metals), while
(b) shows the effect of adding a 
mirror metal component
consisting of elements ${\rm C',\ O',\ Ne',\ Si',\ Fe'}$, with 
$\log [n_{\rm A'}/n_{\rm H'}] = \log [n_{\rm A}/n_{\rm H}](solar \ abundance)] + \zeta $,
where dotted lines are (from top to bottom) $\zeta = +1, 0, -1$.
}
\end{figure}

In order to gain some insight,
it is useful to evaluate the cooling rate for the idealized case of a low density
optically thin plasma with ionization dominated by electron impact. In this limit, 
the ionization state and
cooling function depend only on the local temperature.
In Figure 1, we give our results for the cooling function,
$\Lambda_N \equiv  {\cal C}/(n_{e'} n_t)$ [where $n_t \equiv \sum_{A'} n_{A'}$ is the number density of mirror ions].
In Figure 1a
we include only mirror hydrogen and mirror helium,
while in Figure 1b
we examine the effect of having a metal component.
At high temperature, $T \gtrsim 10^{6.5}$ K, radiative cooling is dominated by bremsstrahlung which features a
${\cal C} \propto \sqrt{T}$ behaviour;
at lower temperature, recombination and line emission become the dominant processes. 
The He II, He I, and H I line emission peaks
are evident in the low temperature region in Figure 1a.
Figure 1b shows the effect of adding a 
mirror metal component
consisting of elements ${\rm C',\ O',\ Ne',\ Si',\ Fe'}$, with 
$\log [n_{\rm A'}/n_{\rm H'}] = \log [n_{\rm A}/n_{\rm H}](solar \ abundance)] + \zeta $.
Figure 1b indicates that the cooling rate is relatively insensitive to a halo mirror metal component, so long
as $\zeta \lesssim 0$. This translates to a metal mass fraction of $f_{\rm metal} \lesssim 0.1\%$.

The idealized case of an optically thin plasma, where ionization is dominated by electron impact, 
is certainly pedagogically very useful.
Of course, in
the numerical analysis to follow we include all sources of ionization, that is, both electron impact
and photoionization. This means that the ionization state and cooling function depend not 
only on the halo temperature but can also depend  on the location within the halo.

Halo heating is powered by
ordinary Type II supernovae - made possible due to the kinetic mixing interaction.   
With a core temperature of around 30 MeV, the SN core contains a plasma of electrons, positrons, photons and neutrinos. 
A small kinetic mixing interaction can
generate a mirror particle plasma; the fundamental particle processes are:
plasmon decay to $e' \bar e'$, 
$e\bar e \to e'\bar e'$, $e' \bar e' \to \gamma' \gamma'$ etc. 
The rate at which the core collapse energy is transferred to light mirror particles is estimated to be 
\cite{raffelt,raffelt2} :
\begin{eqnarray}
Q_P = \frac{8 \zeta_3}{9 \pi^3}  \epsilon^2 \alpha^2 \left(
\mu^2_e + \frac{\pi^2 T_{\rm SN}^2
}{3}\right) T_{\rm SN}^3 Q_1
\label{raf1x}
\end{eqnarray}
where $\zeta_3 \simeq 1.202$ is the Riemann zeta function,
$Q_1$ is a factor of order unity, $\mu_e$ is the electron chemical potential, and 
$T_{\rm SN} \sim 30$ MeV is the supernova core temperature.
If $Q_P$ is assumed to be less than the energy loss rate from neutrino emission, 
then a rough upper limit on $\epsilon$ of
around $\epsilon \lesssim  10^{-9}$  arises 
\cite{raffelt}.
Such low values are consistent with laboratory bounds \cite{rubbia}, 
limits from early universe cosmology \cite{review}, and are also compatible with values
($\epsilon \sim 2 \times 10^{-10}$)
suggested from small scale structure considerations (galaxy mass function) \cite{sunny2}.

The mirror particle plasma ($e', \bar e', \gamma', \nu'$)
generated in the core of a Type II supernova
rapidly expands sweeping up mirror baryons in the process.
Part of the core collapse energy will be transported away from the SN vicinity via 
ordinary neutrinos ($\nu$), via mirror photons ($\gamma'$), and mirror neutrinos
($\nu'$). What remains is halo heating in the SN vicinity via particle
interactions, shocks etc.
We label these four components:
\begin{eqnarray}
{\cal H}_{\rm SN} = L_{\rm SN}^{\nu} + L_{\rm SN}^{\gamma'}  + L_{\rm SN}^{\nu'} + L_{\rm SN}^{e'} 
\ .
\end{eqnarray}
The $\nu'$ component is far too weakly interacting 
to provide any significant halo heating; it can be safely ignored
in this dynamics.
In Paper I, and indeed in earlier work \cite{foot13,foot14}, we
focused on the $L_{\rm SN}^{\gamma'}$ component (labelled in Paper I as $L_{\rm SN}$), 
and assumed that it dominated the halo heating.
The dark photons can be absorbed via the photoionization process, which is 
possible if there is a mirror metal component.
(Interactions of dark photons with free mirror electrons are far too weak to 
trap dark photons and heat  the halo.)
However, it was  found in Paper I that the $\gamma'$ component cannot be readily 
absorbed in the halo, even when
a substantial mirror metal fraction was present. The halo
is excessively ionized, severely suppressing photoionization processes.
Paper I concluded that halo cooling generally exceeds the heating from dark radiation
absorption at realistic halo mass densities.

We are thus left to contemplate the $L^{e'}_{\rm SN}$ component,
the heating of the halo that results from collisional processes, shocks etc. 
This component can potentially provide substantial heating of the halo in a 
local region around the SN. Let us briefly expand on
how this heating would occur.
The huge energy input into light mirror sector particles in the core
of a Type II SN would generate an expanding relativistic `fireball'.
This expanding fireball sweeps up the halo mirror
baryons, of number density $n \sim 0.1\ {\rm cm^{-3}}$, in the vicinity of the SN.
This is a complicated system.
The closest baryonic analogue of this system is the  
fireball model of Gamma Ray Bursts (GRB), e.g. \cite{grb1,grb2,grb3,grb4} 
(see also \cite{fireball} for a review and more detailed bibliography).
That model discusses the evolution of such a relativistic fireball with small baryon load,
which is remarkably analogous to the system under consideration (so analogous in fact, that
mirror SN have been suggested as candidates for the central engine that powers GRBs \cite{blin88,zur88}).

The transfer of all the energy of the fireball to the kinetic energy of the baryons
is the eventual outcome of a fireball with a baryonic load \cite{grb3,grb4}. 
If the baryonic load is sufficiently small the baryons will be accelerated to relativistic energies.
Ultimately, this flow is decelerated and part of the kinetic
energy is reconverted into thermal energy which can radiatively cool, producing
the $L^{\gamma'}_{\rm SN}$ component. The remaining component, the kinetic
energy of the mirror baryons (and mirror electrons) is the $L^{e'}_{\rm SN}$ component. 
This is the broad picture.  The details are, of course, very complicated and in practice it is 
very challenging to calculate, with any certainty, the relative portion of energy converted
into radiation versus kinetic energy of the mirror baryons, $L^{\gamma'}_{\rm SN}/L^{e'}_{\rm SN}$. 
Fortunately, for the purposes of this paper understanding these details is not essential.
The quantity of most interest is the total energy transmitted to the halo per (average) SN, $L_{\rm SN}^{e'}$,
and assuming that this quantity does not vary significantly between galaxies, halo dynamics 
becomes very predictive (as we will see).

The $L_{\rm SN}^{e'}$ halo heating component can be easily incorporated within the 
same formalism of Paper I.
If this heating is sufficiently localized around the SN sources, the 
local heating rate at a given location is proportional to the (suitably time averaged) rate of SN in the vicinity.
Any ``graininess'' effects due to the discrete nature of SN, in both time and space,   
will likely be unimportant given the relevant timescale.
[For a Milky Way scale galaxy it is estimated
\cite{review}
that without a substantial heat source the halo would collapse in a few hundred million years, a timescale that
would encompass very many supernovae.]
It follows therefore, that the
local SN sourced halo heating component takes the form:
\begin{eqnarray}
\frac{dH_{\rm SN}^{e'}}{dV} = f(r) R_{\rm SN} L_{\rm SN}^{e'} 
\end{eqnarray}
where $R_{\rm SN}$ is the galaxy's rate of Type II supernovae and $f(r)$ is the normalized
spatial SN distribution ($\int f(r) dV = 1$).

%

To proceed, we shall assume in this analysis that the supernovae distribution is proportional to the stellar
mass density, Eq.(\ref{doc}), so that:
\begin{eqnarray}
\frac{dH_{\rm SN}^{e'}}{dV} = 
\frac{\kappa e^{-r/r_D}}{4\pi r_D^2 r}
\label{Hn}
\end{eqnarray}
where we have introduced the quantity $\kappa \equiv  
R_{\rm SN} L_{\rm SN}^{e'}$, the SN sourced  local halo heating rate.
This quantity is not a universal constant, but galaxy dependent, 
varying strongly on the SN rate, $R_{\rm SN}$. 
Also, although $L_{\rm SN}^{e'}$ is assumed  
constant, one should keep in mind that
different galaxies can presumably have different
average SN properties, in different environments, so that a weak variation 
of $\kappa$ due to a varying galactic $L_{\rm SN}^{e'}$ value is possible. 

Finally, a few more words on the spherically symmetric modelling 
are perhaps in order.
Considering idealized spherically symmetric systems  
simplifies the dynamics considerably, however the real world is 
more complex: Halos around actual galaxies are expected to show
significant departures from spherical symmetry.
If $L_{\rm SN}^{e'}$ dominates the halo heating, and if this
heating is highly localized around SN, then the 
heating of the halo would be sourced in the disk and transmitted to the bulk
of the halo via conduction, convection, and collective halo motion. 
\footnote{
For a strict spherically symmetric system, the collective (bulk) halo velocity
must be zero. More generally, halo motion is expected.
A particular scenario of interest is halo motion in a plane perpendicular to the galactic
disk. (For the Milky Way and Andromeda, this plane is possibly aligned with the plane of satellites,
cf. discussion in \cite{sunny2}.) 
Even a relatively small bulk halo motion
in this plane
of say, $\sim 30$ km/s, would be sufficient to distribute the disk SN heating over a distance
$\sim 3$ kpc from the disk during a 100 Myr timescale.}
Depending on how effective these processes are,
significant temperature and density gradients might arise as one moves in a direction normal to the disk.
Naturally, the
scale of the departures from spherical symmetry are uncertain. 
Despite this serious issue, this author remains
hopeful that
the analysis of idealized spherically symmetric systems can still provide useful
insight into actual disk galaxies.

\subsection{Analytic dark halo density profile}

Before going into the numerical study of the steady state solutions, 
it is useful to use simple analytic arguments to
estimate the density profiles expected given the assumptions made.
The cooling processes, bremsstrahlung, line emission, and mirror electron capture, each involve two particle
collisions and thus have
a cooling rate: ${\cal C} = \Lambda \rho^2 $.
If the halo evolves to the steady state limit, where local heating and cooling rates balance
(${\cal H} = {\cal C}$),
then $\rho = \sqrt{{\cal H}/\Lambda}$ readily follows. 
With $L_{\rm SN}^{e'}$ dominating the heating ${\cal H} \approx dH_{\rm SN}^{e'}/dV$ 
[Eq.(\ref{Hn})], and $\rho = \sqrt{{\cal H}/\Lambda}$ reduces to:
\begin{eqnarray}
\rho(r) =   \sqrt{ \frac{\kappa}{4\pi \Lambda}  } \ \frac{e^{-r/2r_D}}{r_D \sqrt{r}} 
\ .
\label{r1x}
\end{eqnarray}
In this derivation, an implicit assumption is 
that the halo is optically thin so that heating from the reabsorption of 
cooling radiation is negligible compared with the SN sourced heating.
In general, the cooling function $\Lambda$ will have some nontrivial spatial dependence which
can only be determined by solving the full set of equations for the ionization state, temperature, etc.,
which we will do shortly. However, for the purposes of a simple analytic estimate, we can suppose it is a spatial constant,
which it would be in the limit of a low density optically thin 
isothermal halo.

Consider now the halo rotational (circular) 
velocity, which from
Newton's law is:
\begin{eqnarray}
\frac{v_{\rm halo}^2}{r} = \frac{G_N}{r^2} \int_0^r  \rho(r') \ 4\pi r'^2 dr'  
\label{vhalo}
\end{eqnarray}
where $G_N$ is Newton's gravitational constant.
The halo rotation curve resulting from
the density profile, Eq.(\ref{r1x}), is then:
\begin{eqnarray}
v_{\rm halo}^2 = 
\frac{G_N}{y} \sqrt{ \frac{4\pi \kappa r_D}{\Lambda} } 
\left\{ 3\sqrt{2\pi} \ {\rm erf}(\sqrt{y/2}) - 2e^{-y/2} \sqrt{y} (y+3) \right\}
\end{eqnarray}
where $y \equiv r/r_D$. 
It is useful to define the halo rotation velocity, normalized at a fixed
$r/r_D$ value, conveniently taken to be $r_{\rm opt} = 3.2r_D$:
$v_{\rm halo}(r)/v_{\rm halo}(r=r_{\rm opt})$.
The normalized halo rotation curve is independent of the parameters
$\kappa/\Lambda, \ r_D$, i.e. depends only on the dimensionless ratio $r/r_D$,
and is shown in Figure 2.

\begin{figure}[t]
\centering
\includegraphics[width=0.48\linewidth,angle=270]{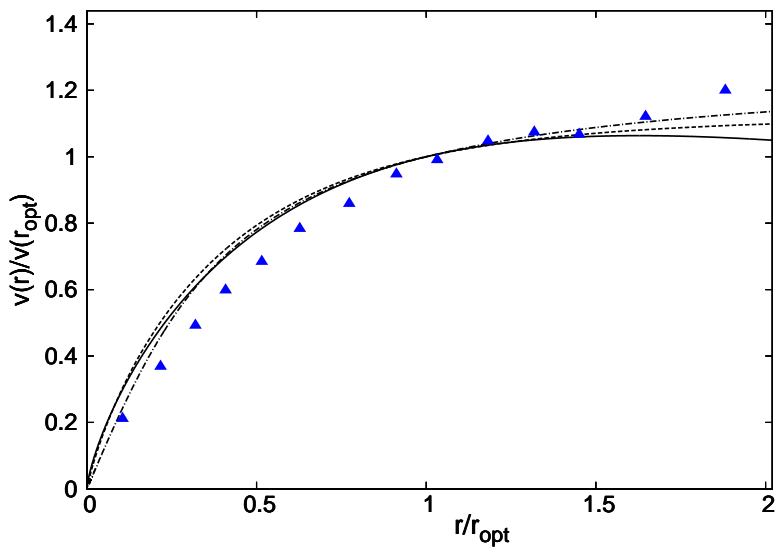}
\caption{
\small
Normalized halo rotational velocity, $v_{\rm halo}(r)/v_{\rm halo}(r=r_{\rm opt})$, $r_{\rm opt} = 3.2r_D$,
resulting from Eq.(\ref{r1x}) [solid line],
which assumes that
that $L_{\rm SN}^{e'}$ dominates halo heating. The dashed line arises from
Eq.(\ref{r1yy}), where  $L_{\rm SN}^{\gamma'}$ is assumed to dominate halo heating.
The dashed-dotted line is the halo rotation curve arising from a quasi-isothermal profile.
Triangles are the synthetic rotation curve derived from 
observations of dwarf galaxies \cite{sal17}. 
}
\end{figure}

In Paper I, in which
dark radiation was assumed to dominate the halo heating, the above simple analytic argument 
applied to radiation heating gave 
a density profile of a somewhat different form. 
In that case $H(r) \propto \rho(r) F(r)$, where $F(r)$ is the flux
of SN sourced dark photons, and  equating local heating with cooling 
suggests $\rho (r) \propto F(r)/\Lambda$. 
The flux, $F(r)$, can be found by integrating over all sources, and
in the
optically thin limit this yields a halo density of the form:
\begin{eqnarray}
\rho(r) = \lambda \int \int \frac{d{\cal L}^S_{\rm SN}(r')}{dV} \ 
{\cal F}(r,r',\theta') \ 2\pi r'^2 d\cos\theta' dr'
\ .
\label{r1yy}
\end{eqnarray}
Here ${\cal F}(r,r',\theta') = 
1/(4\pi [r^2 + r'^2 - 2rr'\cos\theta'])$ and $d{\cal L}^S_{\rm SN}/dV$ is the source luminosity of 
dark photons generated around SN, and has the same spatial form as Eq.(\ref{Hn}).
The coefficient $\lambda$ combines the various proportionality coefficients, which,
in a first order approximation, can be taken as spatially constant.
The normalized halo rotational velocity arising from this profile is independent of $\lambda, r_D$, 
and is also shown in Figure 2 for comparison.
These rotation curves are very similar to each other, and also from the rotation curve 
that arises from the cored quasi-isothermal
profile. The quasi-isothermal profile is a phenomenological profile often employed in the literature: \cite{kent}, and is given by 
\begin{eqnarray}
\rho(r) = \frac{\rho_0 r_0^2}{r_0^2 + r^2}
\ .
\label{isis}
\end{eqnarray}
The dash-dotted line 
in Figure 2 is the rotation curve that results from the quasi-isothermal profile with core radius $r_0 = r_D$.

In this dynamics, the baryonic scale length parameter, $r_D$, plays a central role.
The halo evolves towards a steady state configuration, which is strongly influenced
by the distribution of supernovae, as these represent the primary source of halo heating.
This approach offers a very simple explanation for a dark halo scale length 
that correlates with the baryonic scale length, e.g.
\cite{DS}.
It also has the potential to  explain  the observed rather soft ($\sim$  cored) density profile 
in the inner region ($r \lesssim r_D$)
of rotationally supported galaxies, 
e.g. \cite{154,blok3,littlethings}. 
 
There are a few more things we can do analytically with the density profile, Eq.(\ref{r1x}).
The halo rotation curve has a maximum, which, for a spatially constant $\Lambda$, 
arises at $r \simeq 5.2r_D$, and is given by:
\begin{eqnarray}
\left[ v_{\rm halo}^{\rm max} \right]^2 \simeq 3.12 \ G_{N}  \sqrt{ \frac{\kappa  r_D}{\Lambda}}
\ .
\label{ss1}
\end{eqnarray}
Also note that 
the density profile, Eq.(\ref{r1x}),  leads to a finite halo mass, $m_{\rm halo} = \int \rho dV$:
\begin{eqnarray}
m_{\rm halo} &=& \frac{\sqrt{4\pi \kappa }}{\sqrt{\Lambda}\ r_D} \int e^{-r/2r_D} r^{3/2} dr
\nonumber \\
         &= & 6\pi r_D^{3/2} \ \sqrt{\frac{2\kappa }{\Lambda}} 
\ .
\label{ss2}
\end{eqnarray}
Eq.(\ref{ss2}) relates the halo mass to  
the SN rate and baryonic scale length in a given galaxy.
Such a relation need not require any `fine tuning' of parameters.
The relation should be viewed as the product of dynamical evolution, 
where the baryonic distribution (SN rate and
scale length) has evolved in time in a way that leads to the steady state
configuration, and hence also Eq.(\ref{ss2}).
We will say a few more words on this evolution later on.

Combining Eq.(\ref{ss1}) with Eq.(\ref{ss2}) we find:
\begin{eqnarray}
m_{\rm halo} &\simeq & \frac{8.54}{G_N} \left[ v_{\rm halo}^{\rm max} \right]^2 r_D \nonumber \\
& \simeq & 4.0\times 10^{11} m_\odot \ \left[ \frac{v_{\rm halo}^{\rm max} }{200 \ {\rm km/s}}\right]^2 \left[ \frac{r_D}{5 \ {\rm kpc}}\right]
\label{13xx}
\ .
\end{eqnarray}
Evidently, for a Milky Way scale galaxy with baryonic 
parameters: $m_{\rm baryon} = 10^{11} \ m_\odot$, $r_D \simeq 4.6$ kpc,
Eq.(\ref{13xx}) indicates
a halo with mass around four times that of the baryons.
This is not so far from the observed cosmic abundance derived from CMB analysis, e.g.\cite{planck}.
\footnote{
For the case considered in Paper I, where SN sourced dark photons heat the halo,
no such relation could be derived 
as the density in that case, Eq.(\ref{r1yy}), features a divergent halo mass.  
}
Naturally, though, complex galaxy evolutionary processes can lead to baryon mass 
loss and potentially also halo mass loss, 
so deviations from the cosmic abundance are anticipated. 
Note that one can make use of known approximate scaling relations \cite{btfnew} and \cite{wu} 
to infer that Eq.(\ref{13xx}) implies that $m_{\rm halo}/m_{\rm baryon}$
increases for smaller galaxies: $m_{\rm halo} \propto (m_{\rm baryon})^{0.9}$.

\subsection{The steady state solutions}

The hydrostatic equilibrium condition in Eq.(\ref{SSX}) relates the pressure 
gradient to the gravitational
force. With the assumption that all particle species, ${\rm e', \ H', \ He'}$, are in local
thermodynamic equilibrium at a common temperature T, the fluid pressure is
$P = \rho T/\bar m$. Here  $\bar m \equiv \sum n_i m_i/\sum n_i$ 
is the mean mass of the particles forming the plasma.
In the absence of dark magnetic fields, or bulk halo motion,
the quantities, $\rho, T, P$, or equivalently, $\rho, T, \bar m$, 
fully describe the physical properties of
the fluid. They are each location dependent, and can be determined in the steady state limit by 
solving Eq.(\ref{SSX}) together with  a set of equations 
describing the ionization state (given in Paper I). 

To compute the steady state solutions we follow the same procedure as in Paper I.
At each location in the halo, the hydrostatic equilibrium condition, the ionization state, 
heating and cooling rates, are evaluated. 
Because of the interdependence of these quantities, an iterative procedure 
is followed.
Dark radiation sources, line emission and recombination,
are included as in Paper I, but the halo heating rate is modified to include the $L_{\rm SN}^{e'}$
contribution.  To get a physically interesting  solution for a Milky Way scale galaxy, we will 
require $L_{\rm SN}^{e'}$ to be substantial and to dominate
the energy input into the halo.  In this parameter regime, where 
$L_{\rm SN}^{\gamma'}$ heating is relatively unimportant,
the dark photons could still influence the heating and cooling indirectly.
Any substantial flux of dark photons with energies near the mirror helium
ionization energy could modify the ionization state and potentially lead to important
effects for small galaxies. 
For the purposes of this paper, we shall assume such 
effects are negligible; an assumption that is valid for a wide range of parameters.
\footnote{
Modelling the SN sourced dark photons as a thermal distribution, 
we have checked numerically that our results are insensitive to the effects
of the SN sourced dark photons for a wide range of parameters, including:
$R_{\rm SN} L_{\rm SN}^{\gamma'} \lesssim  10^{44}[\kappa/\kappa_{\rm MW}] $ erg/s, and $T_{\rm SN} \gtrsim 1$ keV.
}

The iterative procedure developed in Paper I requires a suitable parameterization 
for the halo density 
profile $\rho(r)$.  With the density profile parameterized, the temperature profile can be 
derived from the hydrostatic equilibrium condition, 
and the ionization state, local heating and cooling  rates (${\cal H}, {\cal C}$), 
iteratively determined.
(Technically, the temperature profile is also iteratively determined as it depends 
on the ionization state via its dependence on the mean mass parameter, $\bar m$.)
The a priori unknown parameters defining the density profile can 
then be identified by requiring that 
${\cal H} = {\cal C}$.
Although in principle any (sufficiently general) test  function can be used, 
the problem will be numerically simpler to solve if
a suitable function with only a few parameters can be found. The obvious choice is 
motivated by the simple analytic arguments
leading to Eq.(\ref{r1x}), and we take:
\begin{eqnarray}
\rho(r) = {\lambda} \ \frac{e^{-r/2r_D}}{\sqrt{r/r_D}} 
\ .
\label{r1}
\end{eqnarray}
Spatial dependence of $\lambda$ can be accommodated via an expansion:
\begin{eqnarray}
\lambda \to \lambda \left[1 \ + \ \sum_{n=1}^{N} a_n \left(\frac{r}{r_D}\right)^n 
 +  b_n  \left(\frac{r_D}{r}\right)^n\right]
\ .
\label{2grl}
\end{eqnarray}

For $\rho(r)$ to be an approximate steady state solution requires  
${\cal H} \simeq {\cal C}$ at every location in the halo.
As in Paper I, we 
quantify this by
introducing the functional:
\begin{eqnarray}
\Delta \equiv \frac{1}{R_2-R_1} \int^{R_2}_{R_1} 
\frac{|{\cal H}(r') - {\cal C}(r')|}{{\cal H}(r')+{\cal C}(r')} \ dr' 
\ .
\label{d1}
\end{eqnarray}
We take $R_1 = 0.15r_D$, $R_2 = 7.4r_D$ in our numerical work.
The $\Delta$ functional is then minimized with respect to variations of the 
density parameters,
$\lambda, a_n, b_n$.
The density solution is defined by the parameters,  $\lambda, a_n, b_n$,
for which $\Delta [\lambda, a_n, b_n]$ is minimized.
It turns out that keeping only the first few terms [$N=2$ in Eq.(\ref{2grl})]
is sufficient to yield $\Delta_{min} \lesssim 0.01$.
This appears to uniquely determine the
density profile, as only one solution is found for each set of baryonic parameters. 

In the scenario considered in this paper,
with $L_{\rm SN}^{e'}$ dominating the halo heating, we do not require any 
halo metal content, and for simplicity, we shall assume in this analysis that it is negligible
i.e. $f_{\rm metal}\lesssim 0.1\%$ (as discussed earlier).
The possibility that there exists significant halo metal content is, of course, extremely
interesting, and will be examined in a separate paper, that will also
explore the implications of having a dark sector comprised of $N \ge 1$ dark SM copies.
Anyway, with negligible metal content, and with the mirror helium/hydrogen ratio
set by the mirror BBN value (as discussed earlier), the local cooling rate can be calculated within the model.
There are no free parameters given that all the cross sections are known.
Furthermore, if $L_{\rm SN}^{e'}$ does indeed dominate the halo heating, Eq.(\ref{Hn})
indicates that the heating depends only on two parameters: the baryonic scale length, $r_D$, and 
SN halo heating rate, $\kappa \equiv R_{\rm SN} L_{\rm SN}^{e'}$. 
By construction, our spherical galaxies have their baryonic scale length set by
the disk scale length of the disk galaxies they are supposed to represent; this
leaves  $\kappa$ as the only unknown parameter.

\begin{figure}[t]
\centering
\includegraphics[width=0.48\linewidth,angle=270]{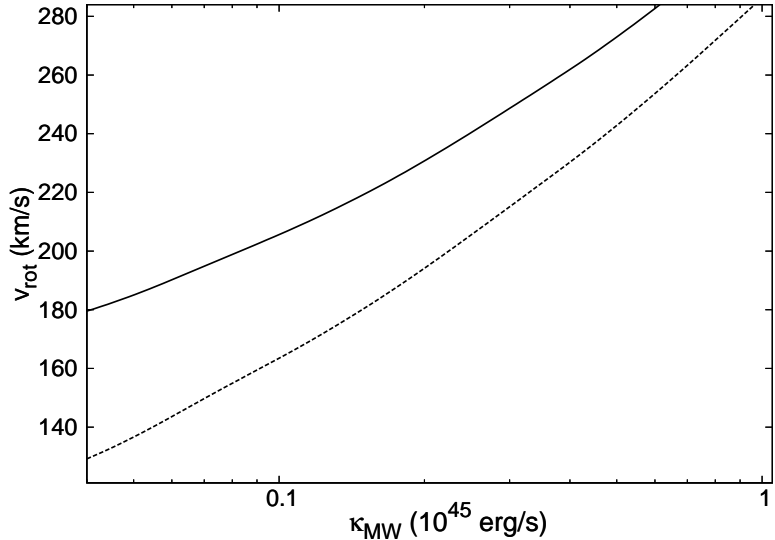}
\caption{
\small
Asymptotic rotational velocity (baryons+halo, solid line) 
and halo velocity (halo only, dashed line)
derived from the
steady state solution for a Milky Way scale
galaxy, with  $m_{\rm baryon} = 10^{11} \ m_\odot$, $f_s = 0.8$, 
$r_D = 4.63$ kpc, and SN heating rate, $\kappa_{\rm MW}$.
}
\end{figure}

Following the procedure described above, 
we have numerically solved the system of equations to yield steady state solutions
for a Milky Way scale galaxy 
for  a variety of $\kappa_{\rm MW}$ values. 
As described above, the
parameters defining the density  [Eq.(\ref{r1}), Eq.(\ref{2grl})]
were ascertained by
minimizing the $\Delta$ functional, Eq.(\ref{d1}), with 
$\Delta_{min} \lesssim 0.01$ obtained in each case.
The rotational velocity, $v_{\rm rot}$, was then determined from Newton's law, 
Eq.(\ref{vhalo}), 
with $\rho \to \rho + \rho_{\rm baryon}$.
To make contact with observations, it will also be convenient to introduce the
asymptotic rotation velocity, which we define as $v_{\rm rot}^{\rm asym} = v_{\rm rot}(r=6.4r_D)$. 

In Figure 3 we give the results for the asymptotic rotation velocity 
derived from the steady state solution in terms of the parameter $\kappa_{\rm MW}$. 
This figure indicates that 
$\kappa_{\rm MW} \approx 2 \times 10^{44}$ erg/s
will be required to achieve a realistic asymptotic rotational  velocity of around 220 km/s for the Milky Way.
This is not far from
the maximum energy available from SN in the Milky Way.
\footnote{The maximum amount of energy transformed into mirror sector particles per SN is of order
the core collapse energy, i.e. $L^{e'}_{\rm SN} \lesssim 3 \times 10^{53}$ erg.
For a Milky Way scale galaxy, the maximum SN rate is around one per decade. 
These estimates suggest an upper limit of
$\kappa_{\rm MW} \lesssim 10^{45}$ erg/s.}

\begin{table}[t]
\centering
\begin{tabular}{c c c c c c}
\hline\hline
$m_{\rm baryon} (m_\odot)$ & $r_D$ (kpc) & $M_{\rm FUV}$ & $f_s$ 
  \\
\hline
${\rm (i)} \ \ 10^{11}$  & 4.63 & -18.4  &     0.8  \\
$\  {\rm (ii)}\ \  10^{10.5}$  & 2.98  &  -17.9 &      0.8 \\
${\rm (iii)}\ 10^{10}$  & 1.91 &   -17.4 &     0.8  \\
${\rm (iv)}\ 10^{9.5}$  & 1.23  &  -16.9 &      0.8  \\
${\rm (v)} \ \ 5 10^{8}$  & 0.60  &  -15.0 &      0.2  \\
\hline\hline
\end{tabular}
\caption{\small Baryonic properties: baryon mass, baryonic scale length, FUV
absolute magnitude, and stellar mass fraction,
for the five `canonical' model galaxies considered.
}
\end{table}


As discussed earlier, $\kappa$ is a galaxy-dependent quantity as $R_{\rm SN}$ varies strongly between different galaxies.
As in Paper I, we assume that the SN rate of a galaxy scales linearly with the galaxy's UV luminosity, $L_{\rm FUV}$.
This is one of the standard measures of a galaxy's current star formation rate, e.g. \cite{uv1,uv2,uv3}.
Provided that $L_{\rm SN}^{e'}$ does not scale significantly between galaxies it follows that
\begin{eqnarray}
\kappa  \approx \kappa_{\rm MW} \ 
\frac{L_{\rm FUV}}{L_{\rm FUV}^{\rm MW}} =  \kappa_{\rm MW}  \ \frac{10^{-0.4M_{\rm FUV}}}{10^{-0.4M_{\rm FUV}^{\rm MW}}}
\label{sc99}
\end{eqnarray}
where $M_{\rm FUV}^{\rm MW} \approx -18.4$ is the FUV absolute magnitude for a Milky Way scale galaxy.
Using the above scaling, and fixing $\kappa_{\rm MW}$ to be $2 \times 10^{44}$ erg/s, there are no parameters left to adjust.
The steady state solution corresponding to any given set of baryonic parameters, $m_{\rm baryon}, f_s, r_D, M_{\rm FUV}$, can
now be evaluated.

\begin{figure}[t]
  \begin{minipage}[b]{0.5\linewidth}
    \centering
    \includegraphics[width=0.7\linewidth,angle=270]{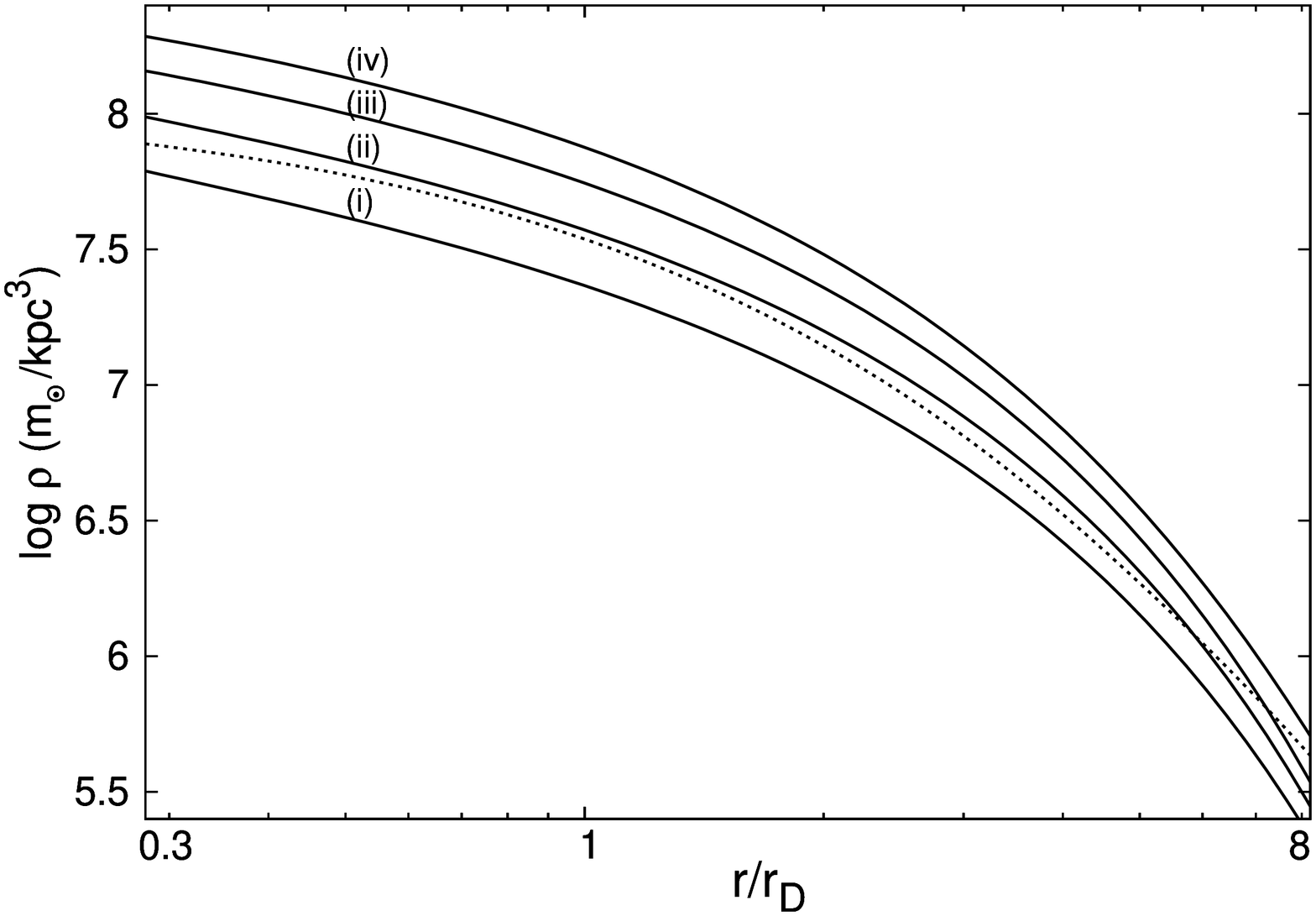}
     (a)
    \vspace{4ex}
  \end{minipage}
  \begin{minipage}[b]{0.5\linewidth}
    \centering
    \includegraphics[width=0.7\linewidth,angle=270]{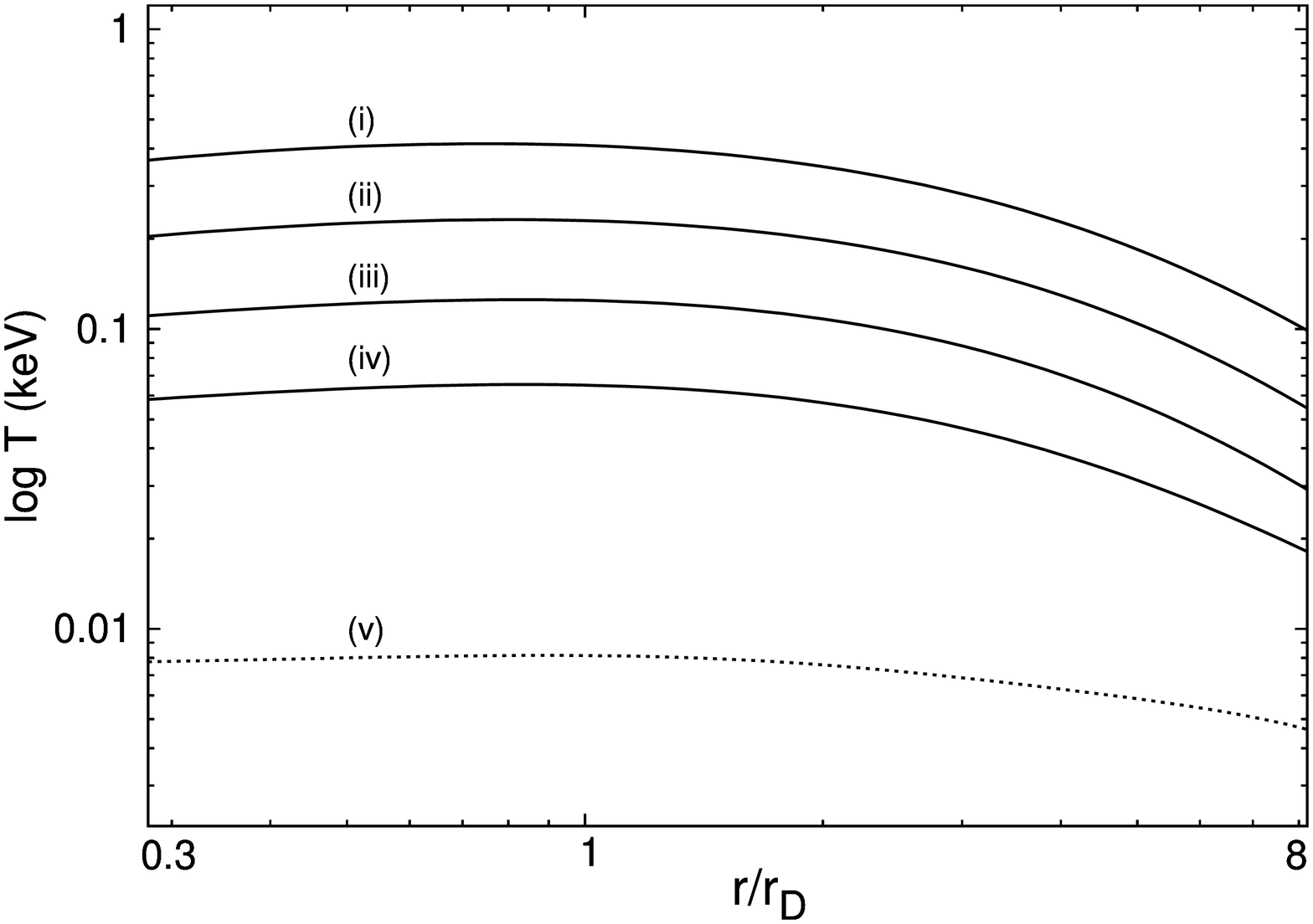}
    (b)
    \vspace{4ex}
  \end{minipage}
  \begin{minipage}[b]{0.5\linewidth}
    \centering
    \includegraphics[width=0.7\linewidth,angle=270]{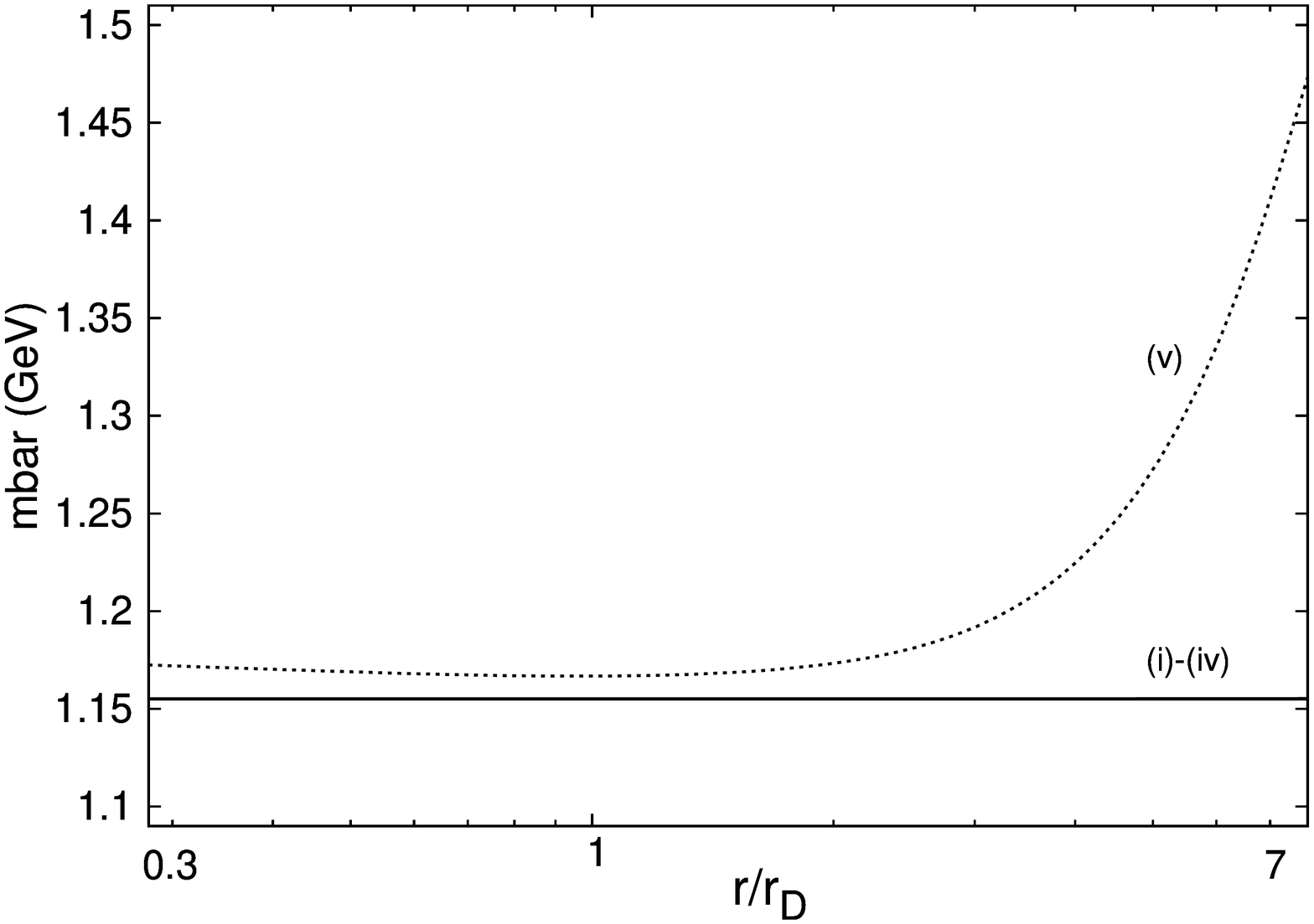}
    (c)
    \vspace{4ex}
  \end{minipage}
  \begin{minipage}[b]{0.5\linewidth}
    \centering
    \includegraphics[width=0.7\linewidth,angle=270]{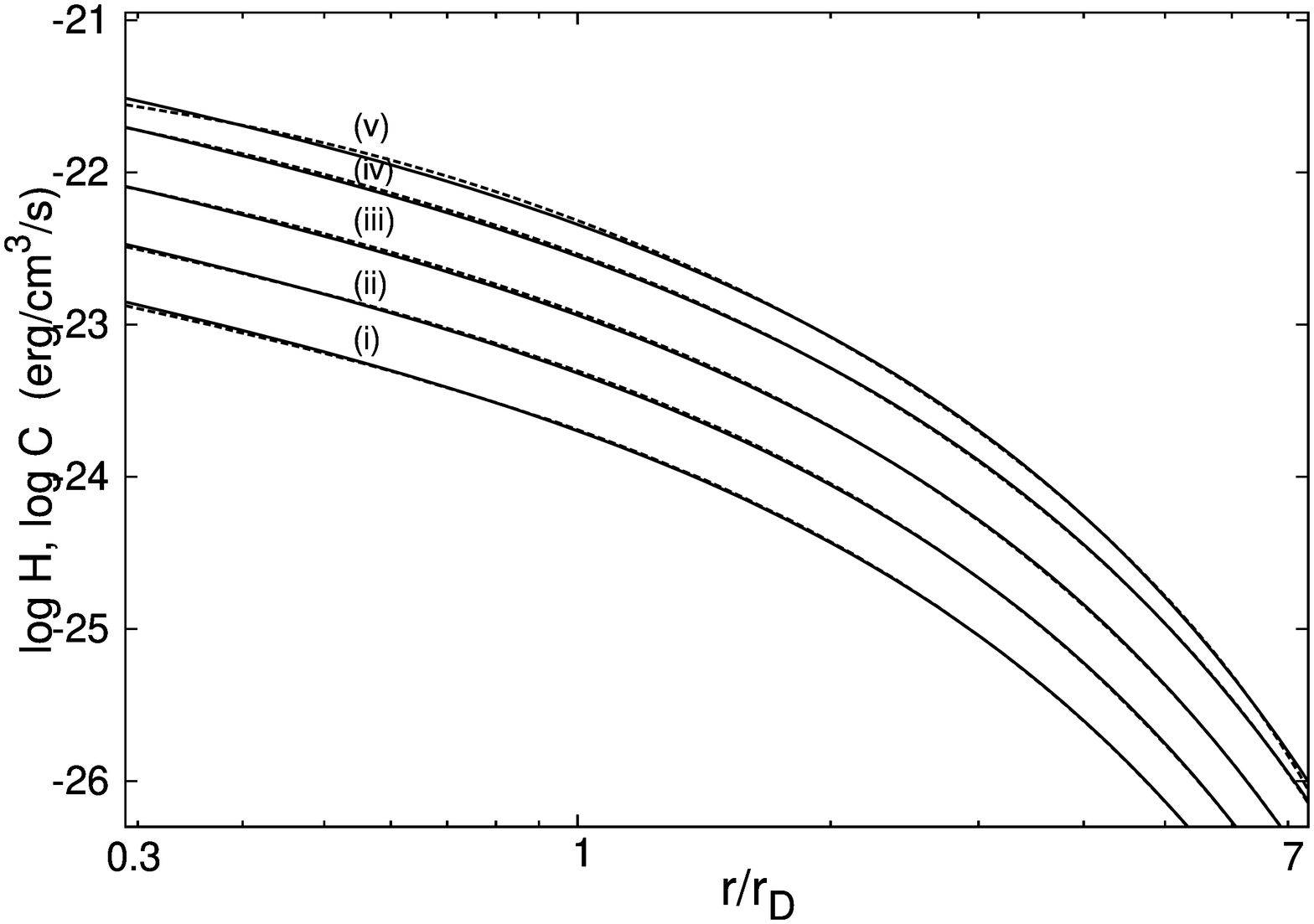}
    (d)
    \vspace{4ex}
  \end{minipage}
\vskip -1.0cm
\caption{\small Properties of the  steady state solutions  computed for 
the set of model galaxies considered.
The baryon mass ranges from
$5\times 10^8 \ m_\odot$ to $10^{11} \ m_\odot$,
see Table 1 for the other baryonic parameters. 
Shown are (a) the
halo density, (b) the halo temperature,
(c) mean mass parameter, $\bar m$,
and (d) the heating, cooling rates [${\cal H}, \ {\cal C}$]  
(solid, dotted line).  
In (a), (b), (c), the dwarf, with $m_{\rm baryon} = 5\times 10^8\ m_\odot$,
is distinguished by a dotted line. 
}
\end{figure}

We consider stellar dominated galaxies  
with baryon masses: $10^{9.5}, \ 10^{10},\ 10^{10.5}$, $10^{11}\ m_\odot$. The  stellar
mass component, with mass fraction  set to $f_s = 0.80$, was
modelled as in Eq.(\ref{doc}). We adopted baryonic scale length ($r_D$)
values matching the stellar disk scale lengths typical of high surface brightness spirals 
(taken from the fit Eq.(1) of \cite{wu}).
The remaining baryon mass fraction ($1-f_s$), the gas component, 
was modelled with a more extended distribution ($r_D^{\rm gas} = 3r_D$, as discussed earlier).
Also considered
was  a gas rich dwarf galaxy (putative dwarf irregular galaxy) 
with baryon mass $5\times 10^8 \ m_\odot$ and  
with stellar mass fraction $f_s = 0.20$.
We adopted
FUV absolute magnitude values compatible with
the measured luminosity \cite{galex} of THINGS \cite{things} and 
LITTLE THINGS \cite{littlethings}
galaxies. 
The galaxy baryonic parameters chosen   
are summarized in Table 1.
For  all of these examples we have numerically solved the system of equations, and thereby 
obtained approximate steady state solutions.

In Figure 4, we give some of the properties of the steady solutions found.
Figure 4a indicates that
the density undergoes significant flattening  at a distance scale $r \sim 2r_D$. 
As discussed earlier, the
flattening of the density profile in the inner region ($r \lesssim 2r_D$) can be understood from the associated
softening of the SN heat source distribution. 

The temperature profiles given in Figure 4b are 
almost isothermal, especially in the region of most interest: $r \lesssim 5r_D$.
There is some softening of the temperature in the outer region and a little softening 
in the inner region, $r \lesssim r_D$.
The figure also shows that
the mean temperature increases for larger galaxies, with a rough scaling: $\langle T \rangle \propto [v_{\rm rot}^{\rm asym}]^2$,
where $v_{\rm rot}^{\rm asym}$ is the asymptotic rotational velocity.
Notice that the temperature region of interest identified from Figure 4b, relevant
for galaxies with active star formation,
$0.007 \lesssim \langle T \rangle/{\rm keV} \lesssim 0.3$, roughly matches the temperature region
between the cooling function dip and ${\rm He \ II}$ line emission peak from Figure 1a.
As will be discussed later on,
this may not be a coincidence as
the negative slope of the cooling function in this `region of interest' might be 
important for ensuring stability of the steady state solution.

The state of ionization 
of the halo can be understood from Figure 4c.
To a good approximation, the halo is fully ionized for the four
largest galaxies considered. In the limit of full ionization,
$n_{e'} = 2n_{\rm He'} + n_{\rm H'}$, and
\begin{eqnarray}
\bar m \simeq  \frac{(n_{\rm H'} + 4n_{\rm He'})m_p}{n_{\rm H'} + n_{\rm He'} + n_{e'}}
        \simeq  \frac{(n_{\rm H'} + 4n_{\rm He'})m_p}{2n_{\rm H'} + 3n_{\rm He'}} 
\end{eqnarray}
where $m_p \simeq 0.94$ GeV is the proton mass. It follows that
 $\bar m \simeq 1.16$ GeV
for the mirror BBN motivated abundance $\log (n_{\rm He'}/n_{\rm H'}) = 0.68$. 
For the dwarf, the halo temperature is low enough so that a significant fraction
of the mirror electrons are bound into mirror helium; the reduction of free
mirror electrons increases $\bar m$ in the regions of low halo temperature.

\subsection{Comparison with the analytic density formula}

In Sec 3.2 an analytic density formula, Eq.(\ref{r1x}), was derived.
Given the simplicity of the analytic formula and the unquestioned usefulness of an analytic
understanding of the behaviour of the numerical solutions, we pause here to study this 
formula in a little more detail. 

The analytic estimate, Eq.(\ref{r1x}), with spatially constant coefficient,
is equivalent to the density, Eq.(\ref{r1}), with spatially constant $\lambda$, and
provides a reasonable first order approximation to the density found in the numerical solutions.
This is shown in Figure 5, which gives the results of a fit 
of the steady state solutions
to the density profile, Eq.(\ref{r1}), with a spatially constant $\lambda$.
The fitted $\lambda \ [10^7 m_\odot/{\rm kpc^3}]$ values for the five canonical galaxies are 
(i) 3.84, (ii) 5.98, (iii) 8.57, (iv) 11.4, and (v) 5.31 .
The derivation of the analytic density formula given in Sec 3.2 assumed that the
halo is optically thin and isothermal; Figure 5 suggests that these conditions
will be  satisfied for the canonical galaxies considered.

\begin{figure}[t]
  \begin{minipage}[b]{0.5\linewidth}
    \centering
    \includegraphics[width=0.7\linewidth,angle=270]{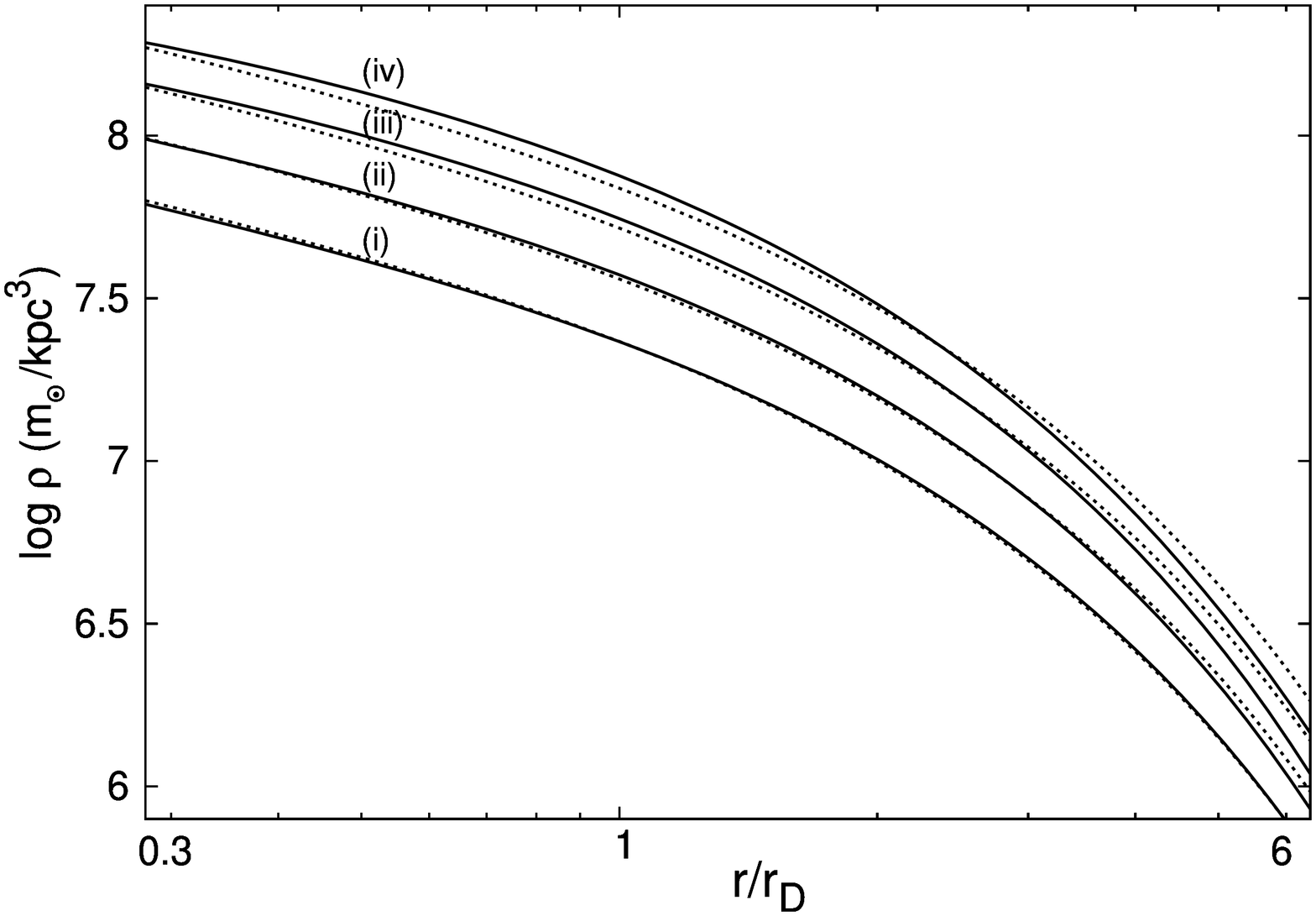}
     (a)
    \vspace{4ex}
  \end{minipage}
  \begin{minipage}[b]{0.5\linewidth}
    \centering
    \includegraphics[width=0.7\linewidth,angle=270]{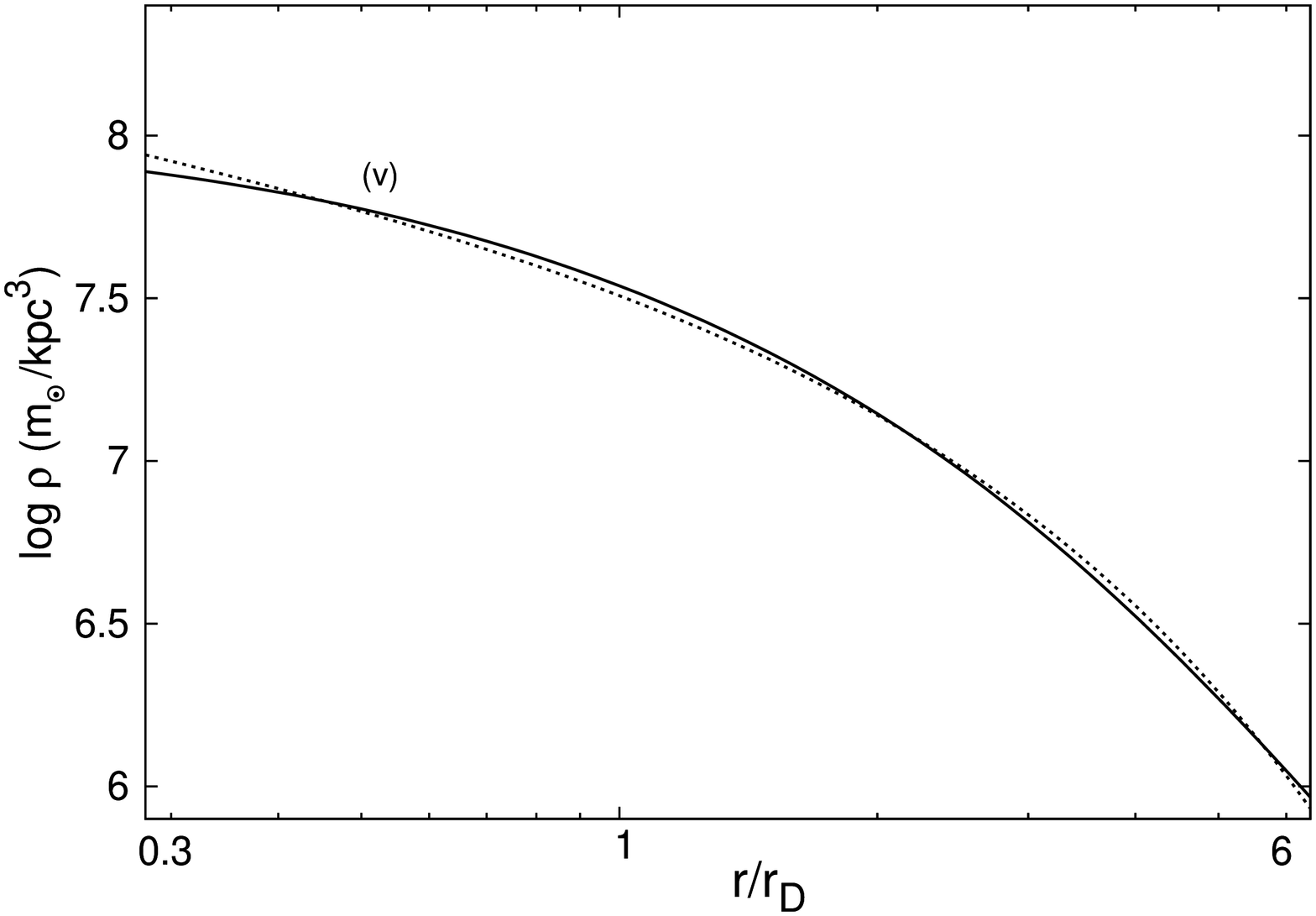}
    (b)
    \vspace{4ex}
  \end{minipage}
\vskip -1.0cm
\caption{
\small 
The steady state density solution (solid line) fitted by the 
density, Eq.(\ref{r1}), with spatially constant $\lambda$ (dotted line).
Panel (a) shows the results for the modelled spirals while panel (b)
is for the modelled dwarf.
}
\end{figure}

We already found (Figure 4b) that the steady state solutions are not too far from  isothermal,
with some departures from isothermality in the outer and inner regions of the halo. 
These corrections are more important for smaller galaxies because their halo temperature is lower (Figure 4b),
where the gradient $\partial \Lambda/\partial T$ 
steepens (Figure 1), resulting in larger effects.
There are also effects as the halo is not always optically thin. 
Of course,
the numerical method for solving the steady state conditions can handle optically thick
regions as reabsorption of cooling radiation is taken into account.
We have calculated (Figure 6) the optical depth for a dark photon that originates at the 
Galactic Center.
Figure 6 indicates that there are some
frequency regions around the He II ionization energy where the halo can be optically thick, especially for 
smaller galaxies; this effect would also contribute to the difference between the analytic density formula and the density
found in the numerical solution.
[Note that for the dwarf, most cooling radiation has photon energy below the He II ionization energy 
which limits the magnitude of the effects of radiation reabsorption for that system.]

\begin{figure}[t]
  \begin{minipage}[b]{0.5\linewidth}
    \centering
    \includegraphics[width=0.7\linewidth,angle=270]{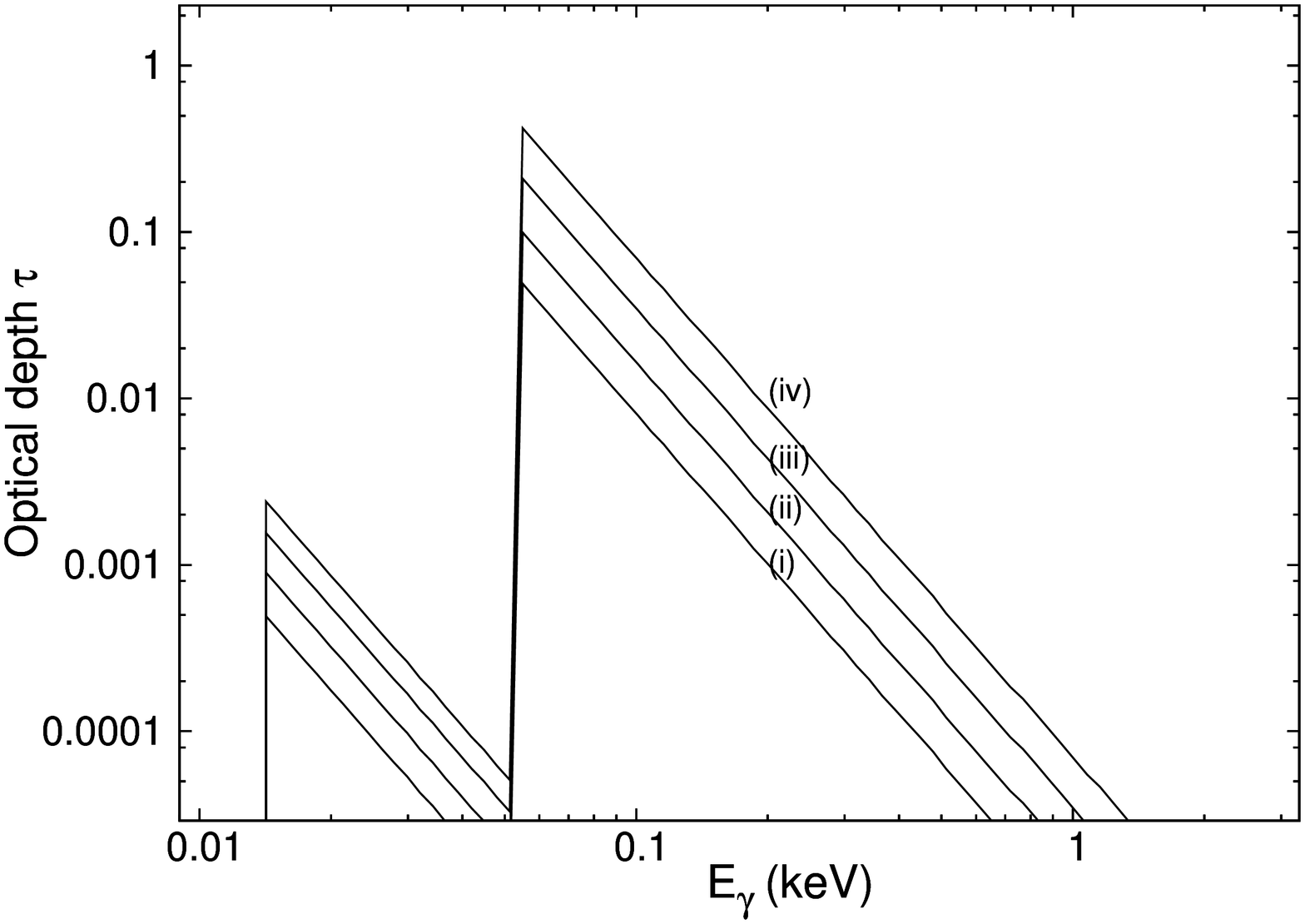}
     (a)
    \vspace{4ex}
  \end{minipage}
  \begin{minipage}[b]{0.5\linewidth}
    \centering
    \includegraphics[width=0.7\linewidth,angle=270]{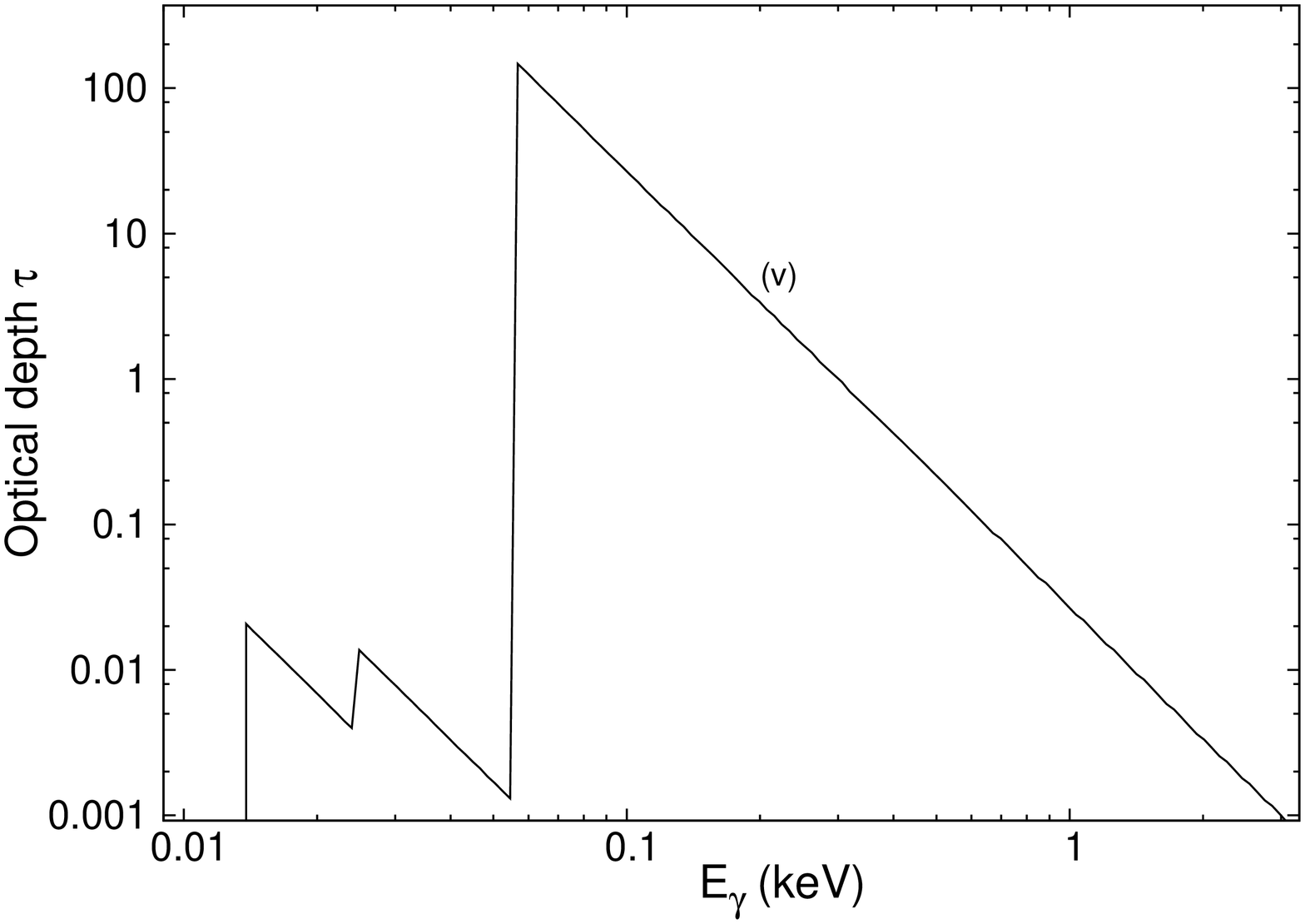}
    (b)
    \vspace{4ex}
  \end{minipage}
\vskip -1.0cm
\caption{
\small
The optical depth for a dark photon originating near the Galactic Center and escaping a galaxy, with
halo properties computed from the steady state solution for each of the canonical galaxies of Table 1.
Panel (a) shows the results for the modelled spirals while panel (b)
is for the modelled dwarf.
}
\end{figure}

To summarize,
the analytic density formula of Sec 3.2, Eq.(\ref{r1x}), along with an isothermal temperature profile, appears 
to provide a reasonable approximation to the physical properties of the halos of the modelled galaxies. 
As discussed above, the origins of 
the small deviation between the analytic formula and the
numerical steady state solutions can be understood in terms of the
modest departures from isothermality and optically thin halo.

\subsection{Comparison with the Burkert Profile}

The halo density that is generated in this dynamics resembles the empirical density profiles
discussed in the literature. Here we shall briefly digress to explore this connection.
Specifically, we consider the quasi-isothermal profile [Eq.(\ref{isis})] and
the Burkert profile \cite{burkert}:
\begin{eqnarray}
\rho (r) &=& \frac{\rho_0 r_0^3}{(r+r_0)(r^2 + r_0^2)} 
\ .
\label{bur}
\end{eqnarray}
We have found numerically that
the Burkert profile provides a somewhat better fit (than the quasi-isothermal profile) 
to the steady state density solution for our canonical galaxies, 
and this fit is shown in Figure 7.
The figure indicates that this profile gives a reasonable accounting of the
computed halo density.  
There are some departures at low radii, $r \lesssim 0.5r_D$, which diminish for smaller galaxies.
That is, for smaller galaxies the Burkert profile improves as an approximation to the computed halo density.
The fitted parameters $\rho_0, \ r_0$ are given in Table 2.

\begin{figure}[t]
  \begin{minipage}[b]{0.5\linewidth}
    \centering
    \includegraphics[width=0.7\linewidth,angle=270]{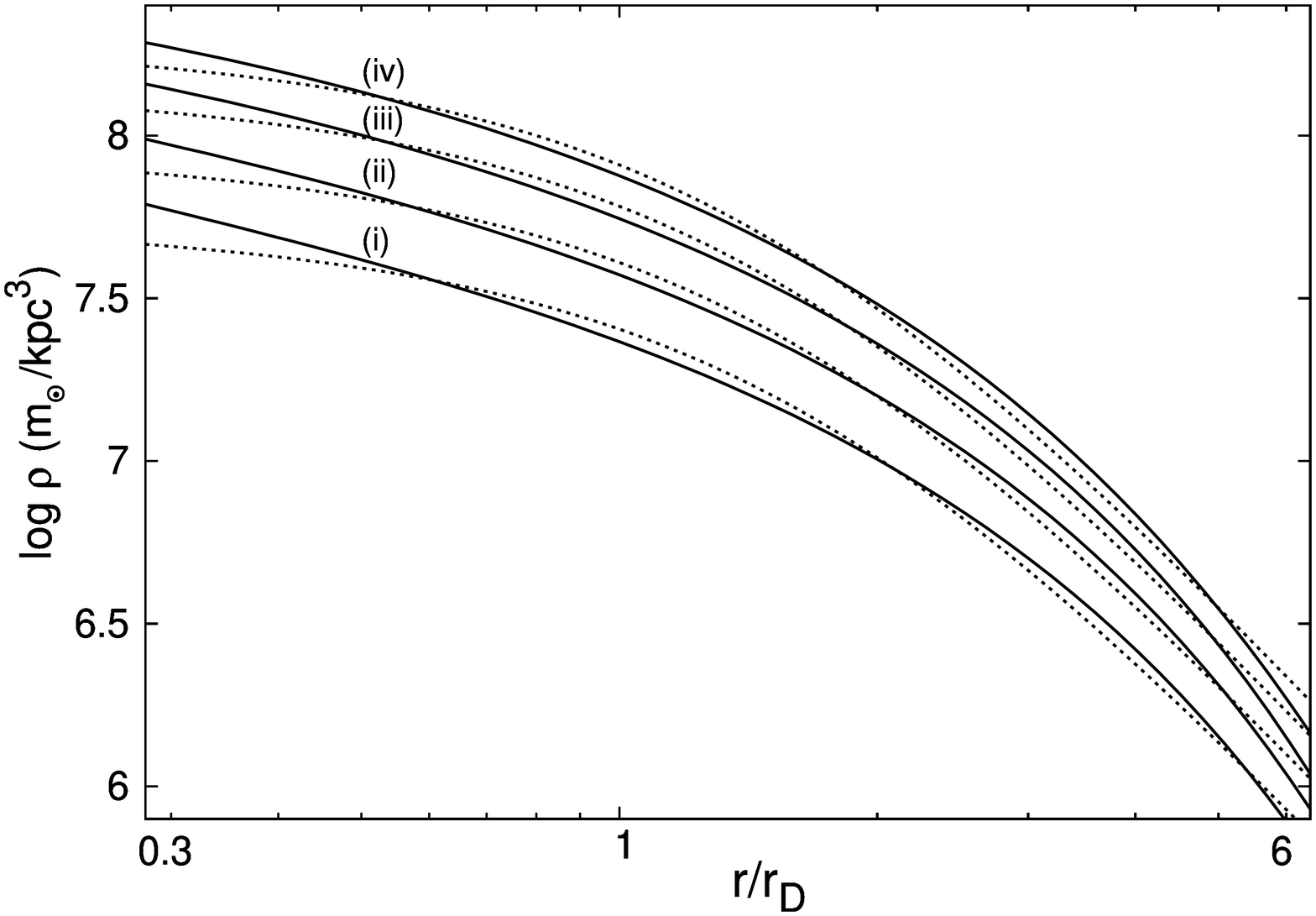}
     (a)
    \vspace{4ex}
  \end{minipage}
  \begin{minipage}[b]{0.5\linewidth}
    \centering
    \includegraphics[width=0.7\linewidth,angle=270]{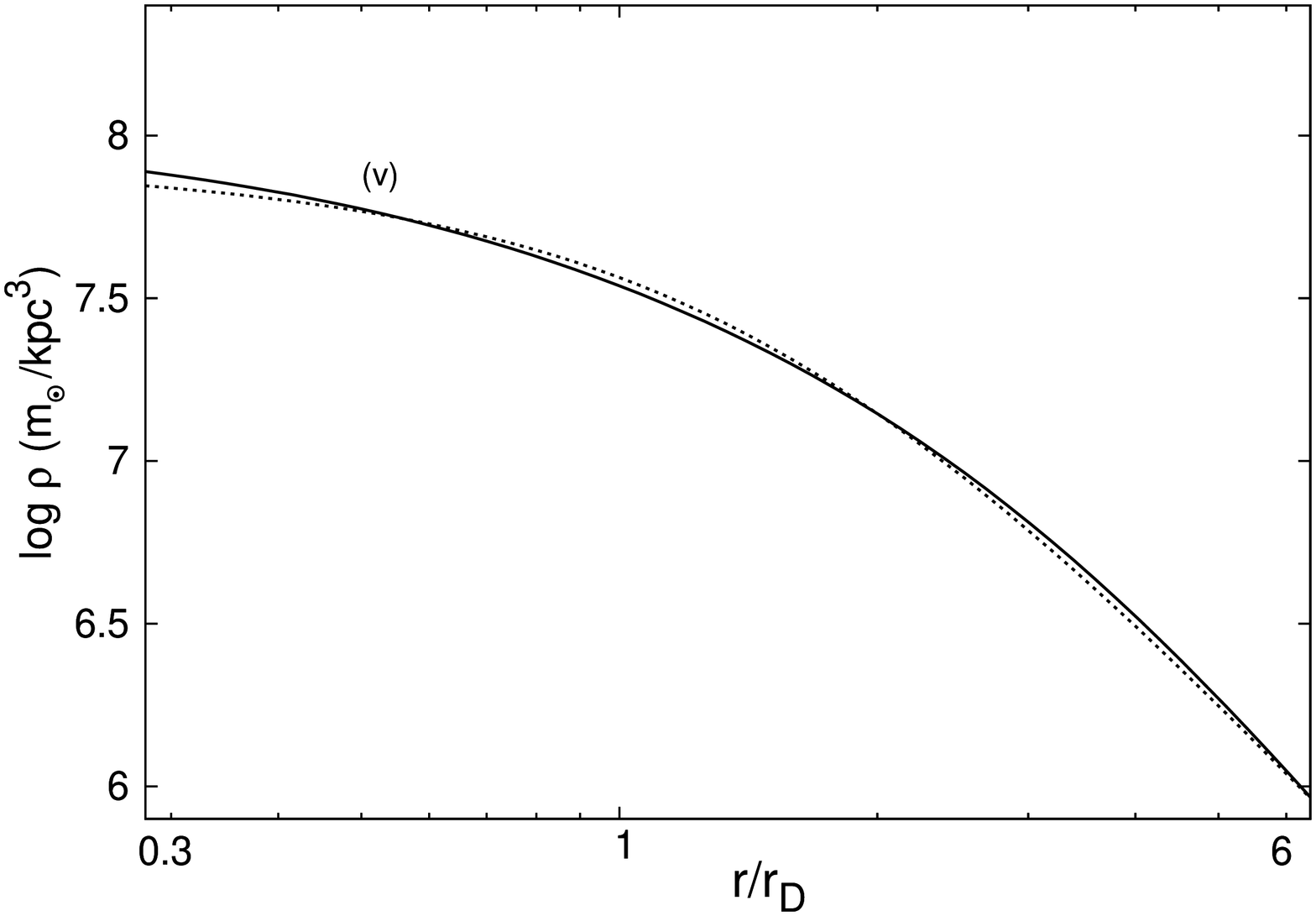}
    (b)
    \vspace{4ex}
  \end{minipage}
\vskip -1.0cm
\caption{
\small 
The steady state density solution (solid line) fitted by the 
Burkert profile, Eq.(\ref{bur})  (dotted line).
Panel (a) shows the results for the modelled spirals while panel (b)
is for the modelled dwarf.
}
\end{figure}

\begin{table}[b]
\centering
\begin{tabular}{c c c c c}
\hline\hline
$m_{\rm baryon} (m_\odot)$ & $r_D$ (kpc) & $r_0/r_D$ & $\rho_0 [10^7 m_\odot/{\rm kpc^3}]$ & ${\rm log}(\rho_0r_0\ [m_\odot/{\rm pc^2}])$ 
\\
\hline
${\rm (i)}\ \ 10^{11}$    & 4.63 & 1.65  & 5.57 & 2.63  \\
$\ {\rm (ii)}\ \ 10^{10.5}$ & 2.98 & 1.58  & 9.33 & 2.64   \\
${\rm (iii)}\ 10^{10}$  & 1.91 & 1.49  &  14.7  & 2.62 \\
${\rm (iv)}\ 10^{9.5}$  & 1.23 & 1.45  &  20.2 & 2.56 \\
${\rm (v)}\ \ 5 10^{8}$  & 0.60 & 1.55  &  8.55 &  1.90  \\
\hline\hline
\end{tabular}
\caption{\small 
Fit of the steady state density solution for the canonical galaxies of Table 1
in terms of the Burkert profile [Eq.(\ref{bur})].
}
\end{table}


Phenomenological fits of rotation curves with these profiles have found that $r_0$ scales
with $r_D$. Indeed, for the isothermal profile ref.\cite{DS} obtained:
\begin{eqnarray}
r_0 = 13 \left(\frac{r_D}{5\ {\rm kpc}}\right)^{1.05}\ {\rm kpc}
\ .
\end{eqnarray}
That is,
$r_0 \approx 2.5 r_D$.
Another empirical scaling relation is that
the dark matter halo surface density, $\rho_0 r_0$, is roughly constant \cite{Donato2,KF},
and for the Burkert Profile is:
\begin{eqnarray}
\log \left(\rho_0 r_0\ [m_\odot/{\rm pc}^2]\right) = 2.15\pm 0.20  
\ .
\label{ds2}
\end{eqnarray}
These relations are broadly compatible with the $\rho_0, \ r_0$  values identified in Table 2.
However there are differences. One of these differences 
is that the $r_0/r_D$ values extracted  from the fit to the steady state solutions
are somewhat lower than values inferred from observations.

One more item of interest is that
the dwarf has a $\rho_0 r_0$ value somewhat below that of the modelled spirals.
As we will discuss in more detail in the following section, this dwarf-spiral discrepancy appears to be confirmed
when the normalization of the halo rotational velocity is compared with the LITTLE THINGS dwarfs.
The origin of these discrepancies is unclear, but naturally the spherically symmetric modelling
of disk galaxies is a likely suspect, especially for the $r_0/r_D$ difference. 

Finally, it is perhaps worth mentioning here that the origin of the two scaling  relations, $r_0 \propto r_D$ and $\rho_0 \propto 1/r_0$
appear quite different. The first one can be understood
from the exponential profile 
of the SN heat sources, which of course is expected to scale with the disk scale length.
It is therefore not a very surprising outcome of this dynamics.
The second relation, on the other hand, appears a little unexpected. From the analytic density
formula, Eq.(\ref{r1x}), one expects the central density to satisfy: $\rho_0 \propto \sqrt{L_{\rm FUV}/\Lambda}/r_D^{3/2} $.
Thus, $\rho_0 r_0$ being constant would
require, roughly, that $L_{\rm FUV} \propto \Lambda r_D$, which appears to be empirically valid for spirals, although its origin is unclear. 

\section{The steady state rotation curves}

\subsection{Dependence on the baryonic parameters}

We now consider the rotation curves in a little more detail.
These curves can be obtained from Eq.(\ref{vhalo}) for the halo 
contribution, and from the same equation with $\rho \to \rho + \rho_{\rm baryon}$
for the full rotation curve.
The rotation curves corresponding to the steady state solutions found are given in Figure 8.
The rotation curves show an almost linear rise 
in the inner region, turning over
to a roughly flat asymptotic profile, with the transition radius occurring at $r \sim 2r_D$
in each case.
As already mentioned, these properties can be understood from
simple geometrical considerations as the halo mass density is closely aligned with the
distribution of supernova sources.
These properties closely resemble the observed rotation curves of spiral,
low surface brightness, and dwarf irregular galaxies,
e.g. \cite{154,blok0,blok1,blok2,blok3,littlethings,salucci,DS,Lelli2,stacy}.

\begin{figure}[t]
  \begin{minipage}[b]{0.5\linewidth}
    \centering
    \includegraphics[width=0.7\linewidth,angle=270]{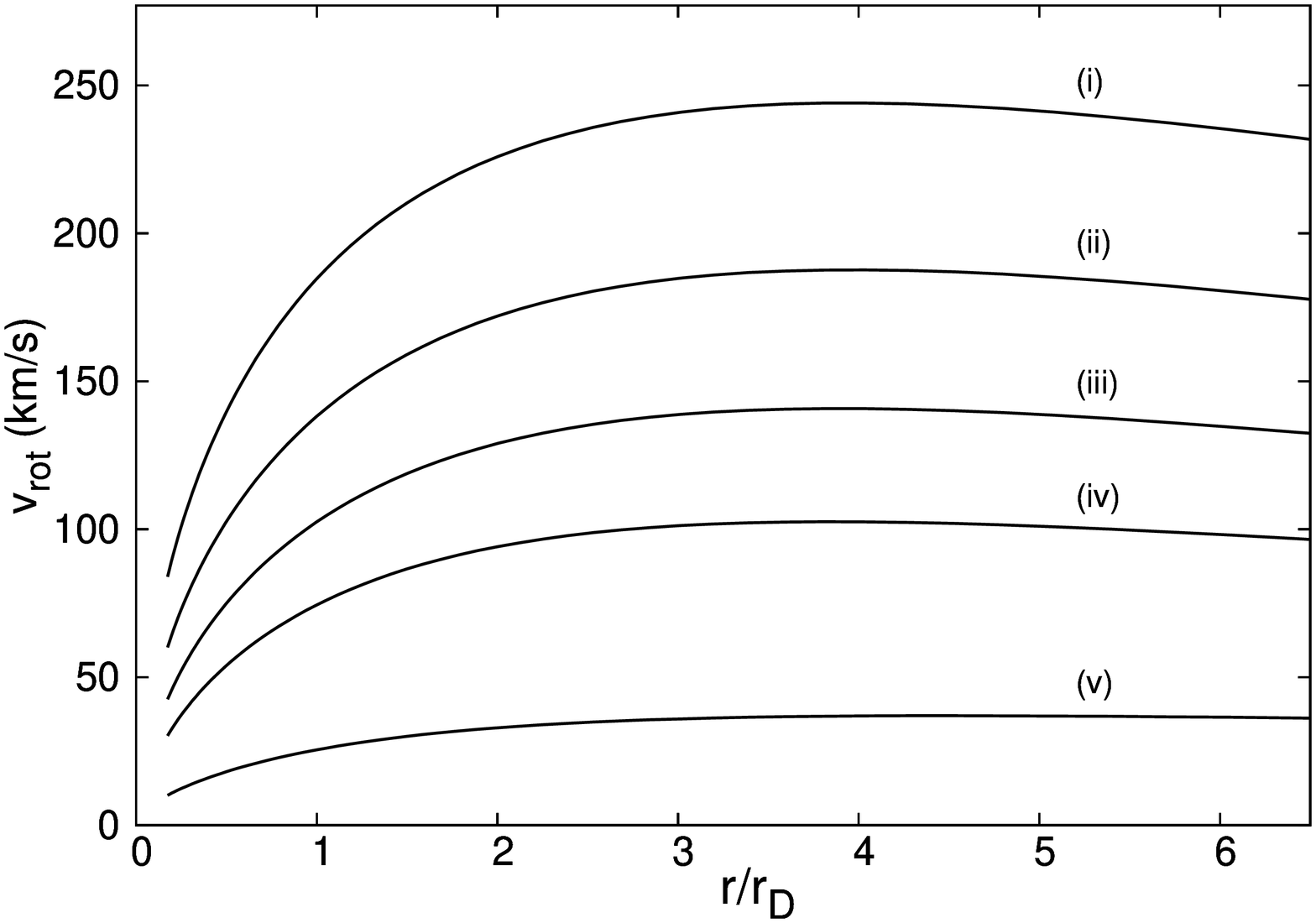}
     (a)
    \vspace{4ex}
  \end{minipage}
  \begin{minipage}[b]{0.5\linewidth}
    \centering
    \includegraphics[width=0.7\linewidth,angle=270]{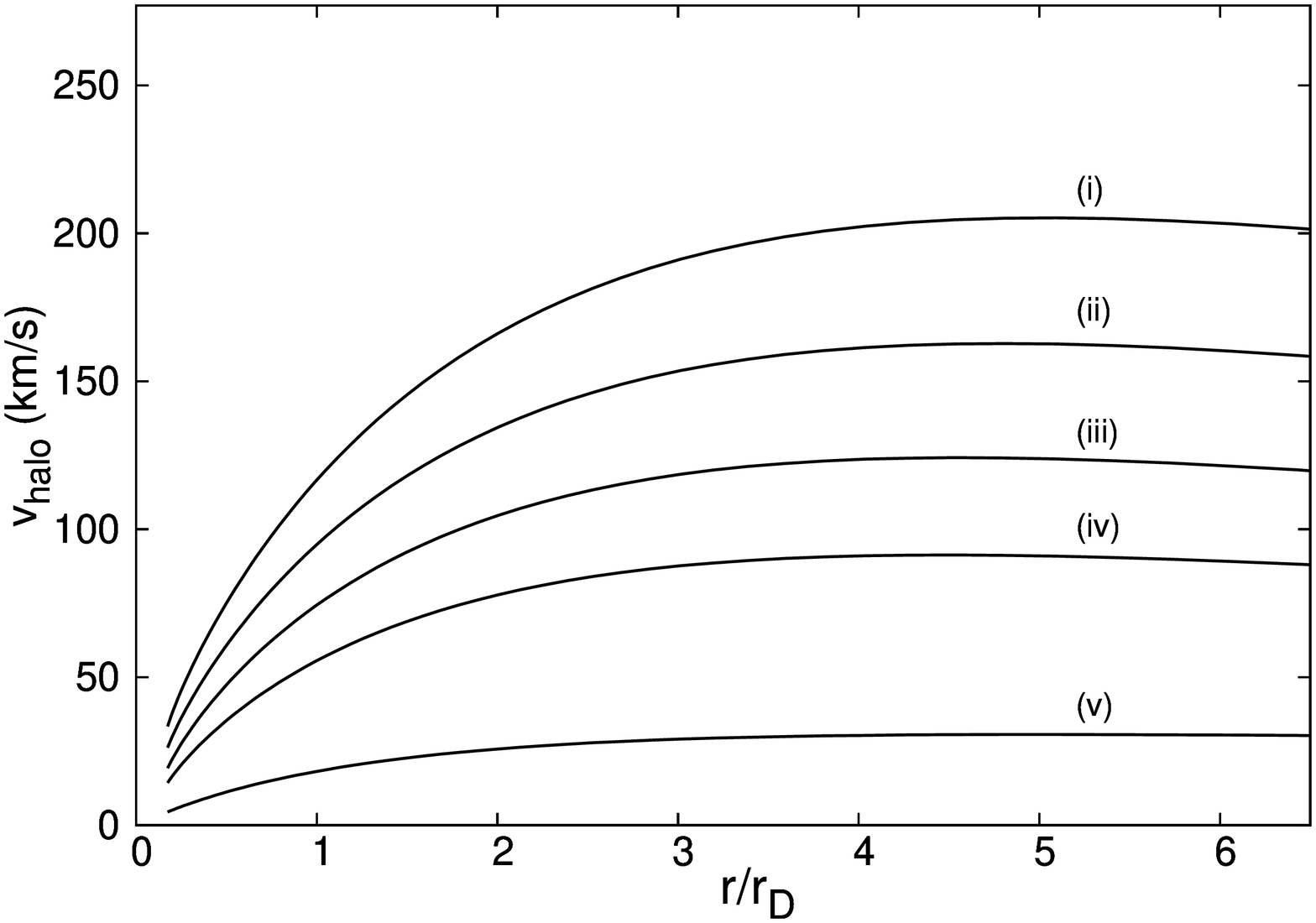}
    (b)
    \vspace{4ex}
  \end{minipage}
\vskip -1.0cm
\caption{
\small 
(a) The rotation curves (halo + baryons)
derived from the computed steady state solutions. 
The baryon mass ranges from
$m_{\rm baryon} = 5\times 10^8 \ m_\odot$ (bottom curve) to $m_{\rm baryon} = 10^{11} \ m_\odot$ (top curve).
See Table 1 for other baryonic parameters chosen. (b)
the corresponding halo rotation curves (halo contribution only).
}
\end{figure}

The steady state solutions depend sensitively on the galaxy's luminosity and baryonic scale length, $L_{\rm FUV}$, $r_D$,  
as these quantities determine the magnitude and distribution
of the SN sourced halo heating.
The dynamics depends weakly on the baryon mass and stellar mass fraction, $m_{\rm baryon},\ f_s$.
These quantities affect the dynamics primarily via their gravitational influence which in turn alters
the temperature profile given the hydrostatic equilibrium condition.
The change in temperature can modify cooling, but such alterations do not 
lead to large effects as the cooling function is fairly flat in the relevant
temperature region for spirals (Figure 1).
Of course in actual galaxies the luminosity is correlated with the baryon mass which
can hamper attempts to check the expected indifferent behaviour.


\begin{figure}[t]
  \begin{minipage}[b]{0.5\linewidth}
    \centering
    \includegraphics[width=0.7\linewidth,angle=270]{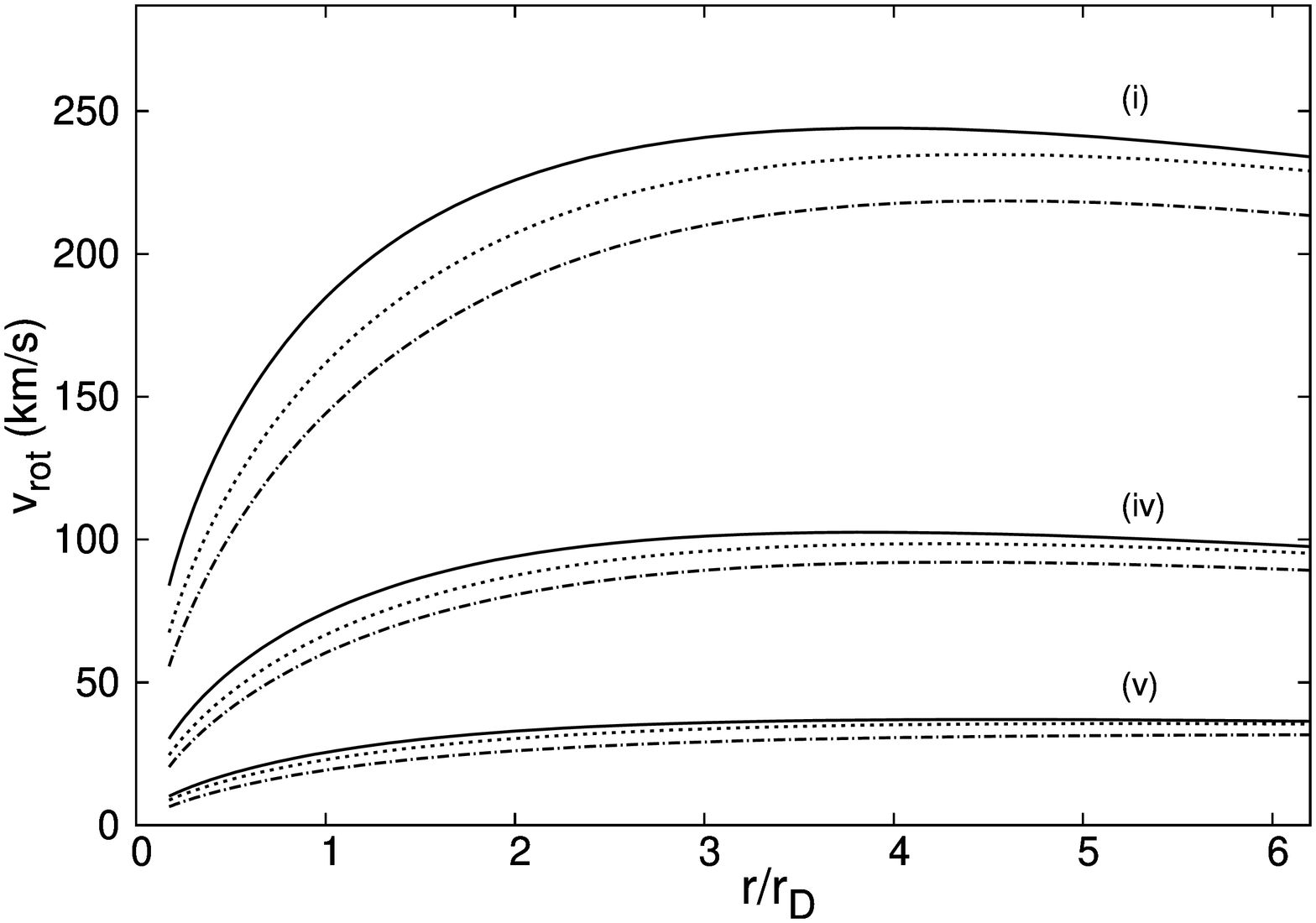}
     (a)
    \vspace{4ex}
  \end{minipage}
  \begin{minipage}[b]{0.5\linewidth}
    \centering
    \includegraphics[width=0.7\linewidth,angle=270]{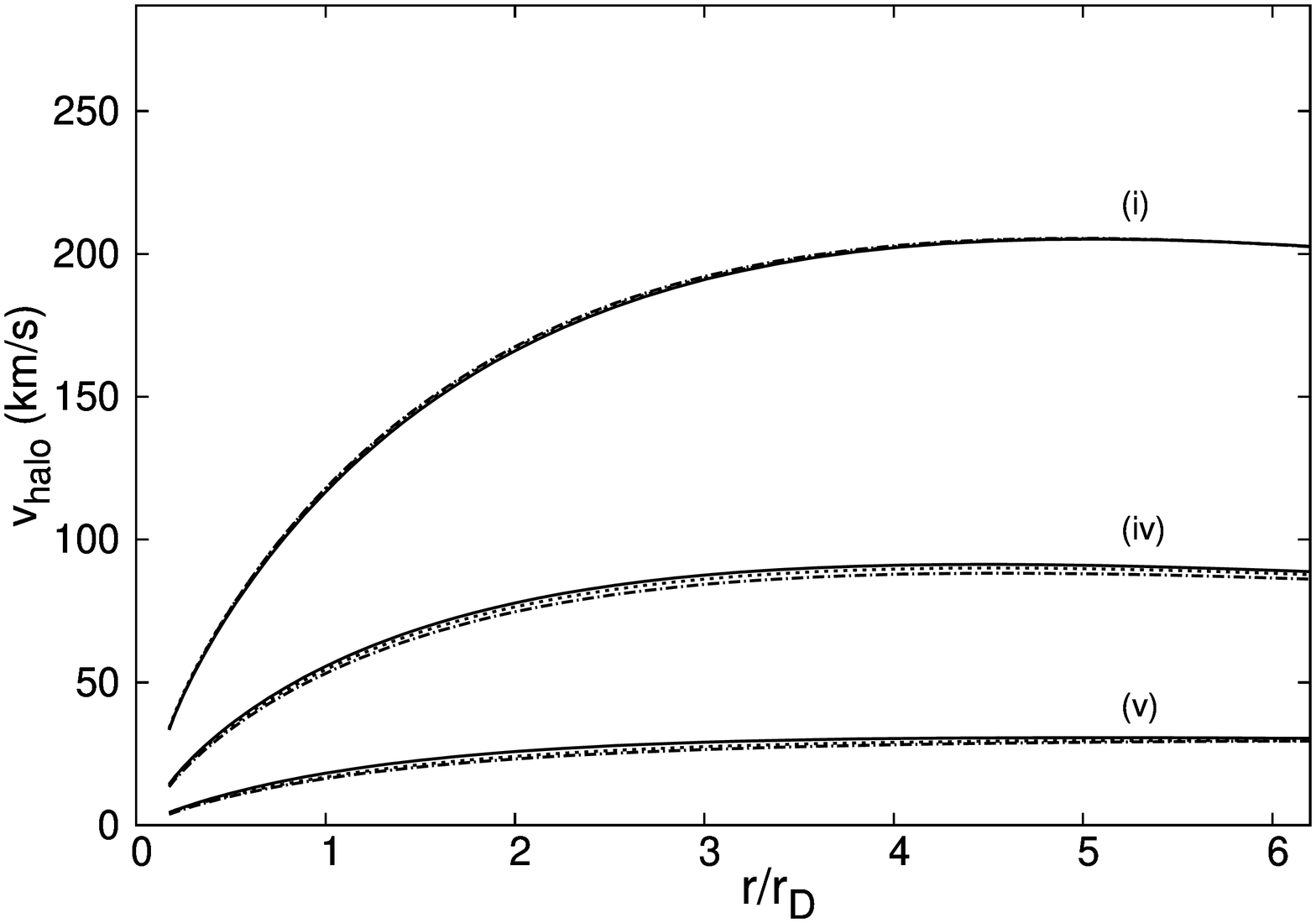}
    (b)
    \vspace{4ex}
  \end{minipage}
\vskip -1.0cm
\caption{
\small
(a) Steady state rotation curves (halo + baryons) for the canonical baryonic parameters (i), (iv), (v) (solid line), 
a $m_{\rm baryon} \to m_{\rm baryon}/3$ parameter variation (dashed dotted line) 
and a $f_s \to f_s/2$ parameter variation (dotted line).
(b) The corresponding halo rotation curves.
}
\end{figure}

The steady state solutions were recomputed allowing for modifications to the baryon mass, and stellar
mass fraction parameters. Specifically we considered galaxies with $m_{\rm baryon} \to m_{\rm baryon}/3$,
with the other baryonic parameters unchanged from Table 1 values. We also considered the variation, $f_s \to f_s/2$, 
again keeping other baryonic parameters fixed.
The effect on the resulting rotation curves of these parameter variations are shown in Figure 9.
The effects are indeed rather minor, consistent with the discussion above.

The baryonic scale length dependence is the next item on the agenda. 
Increasing this parameter, $r_D \to 2r_D$, with the other baryonic parameters fixed to their Table 1 values, we 
again recompute the  steady state solutions.
In Figure 10 we show the resulting rotation curves.
As indicated by this figure, there is some modest $r_D$ dependence on 
the normalization of the halo
rotational velocity with 
no discernible effect on the shape. 
\footnote{
This is a little different to the case
of halo heating dominated by SN sourced dark photons that was considered in  
Paper I.  As discussed there, and in earlier related work \cite{zurab}, 
the halo rotation curves in that case depend approximately only
on the dimensionless parameter, $r/r_D$. 
}
Note that the maximum halo rotational velocity values found in the numerical solutions 
are approximately consistent with: $v_{\rm halo}^{\rm max} \propto r_D^{1/4}$,
the rough analytical result derived in section 3.2 [Eq.(\ref{ss1})].

\begin{figure}[t]
  \begin{minipage}[b]{0.5\linewidth}
    \centering
    \includegraphics[width=0.7\linewidth,angle=270]{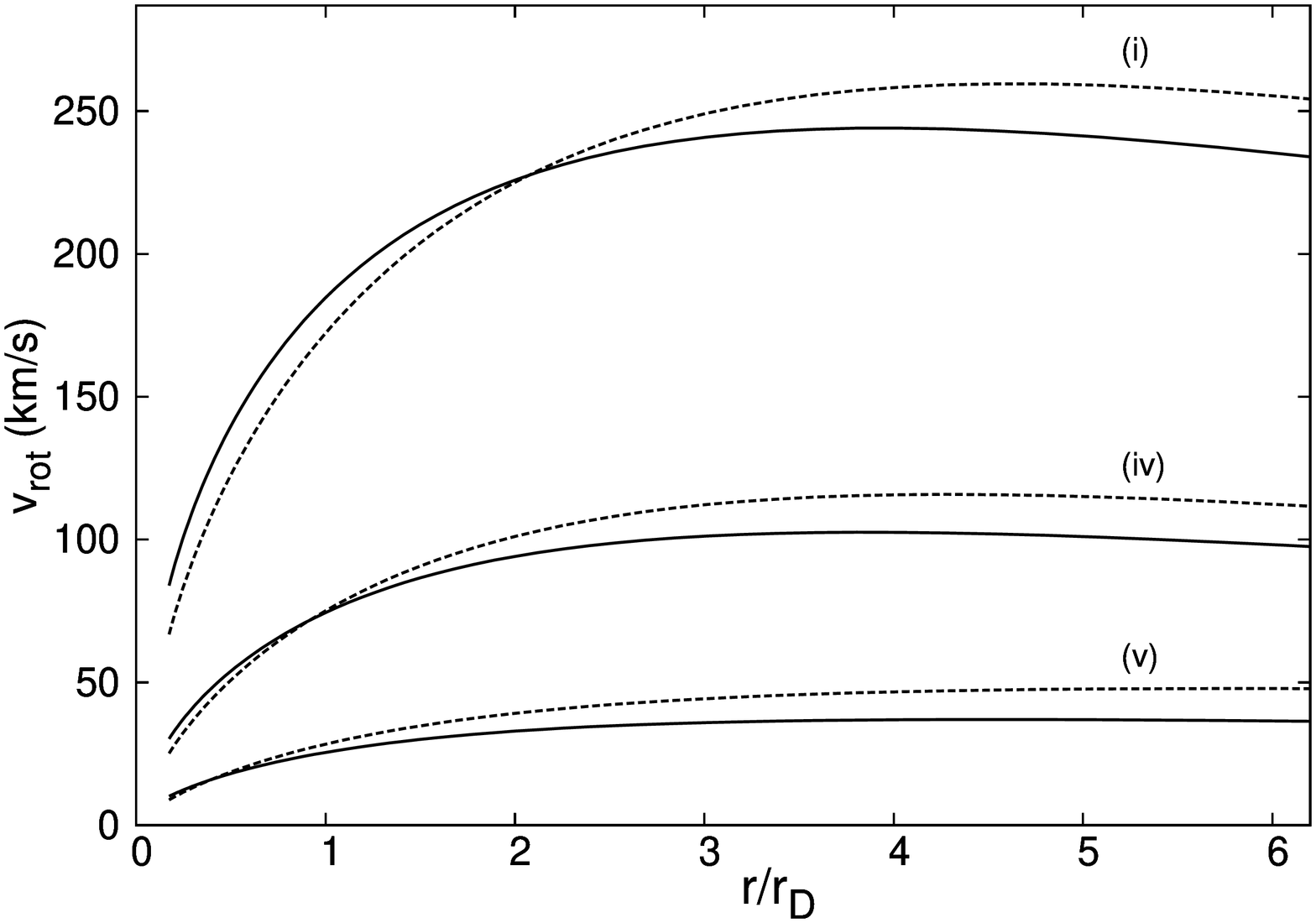}
     (a)
    \vspace{4ex}
  \end{minipage}
  \begin{minipage}[b]{0.5\linewidth}
    \centering
    \includegraphics[width=0.7\linewidth,angle=270]{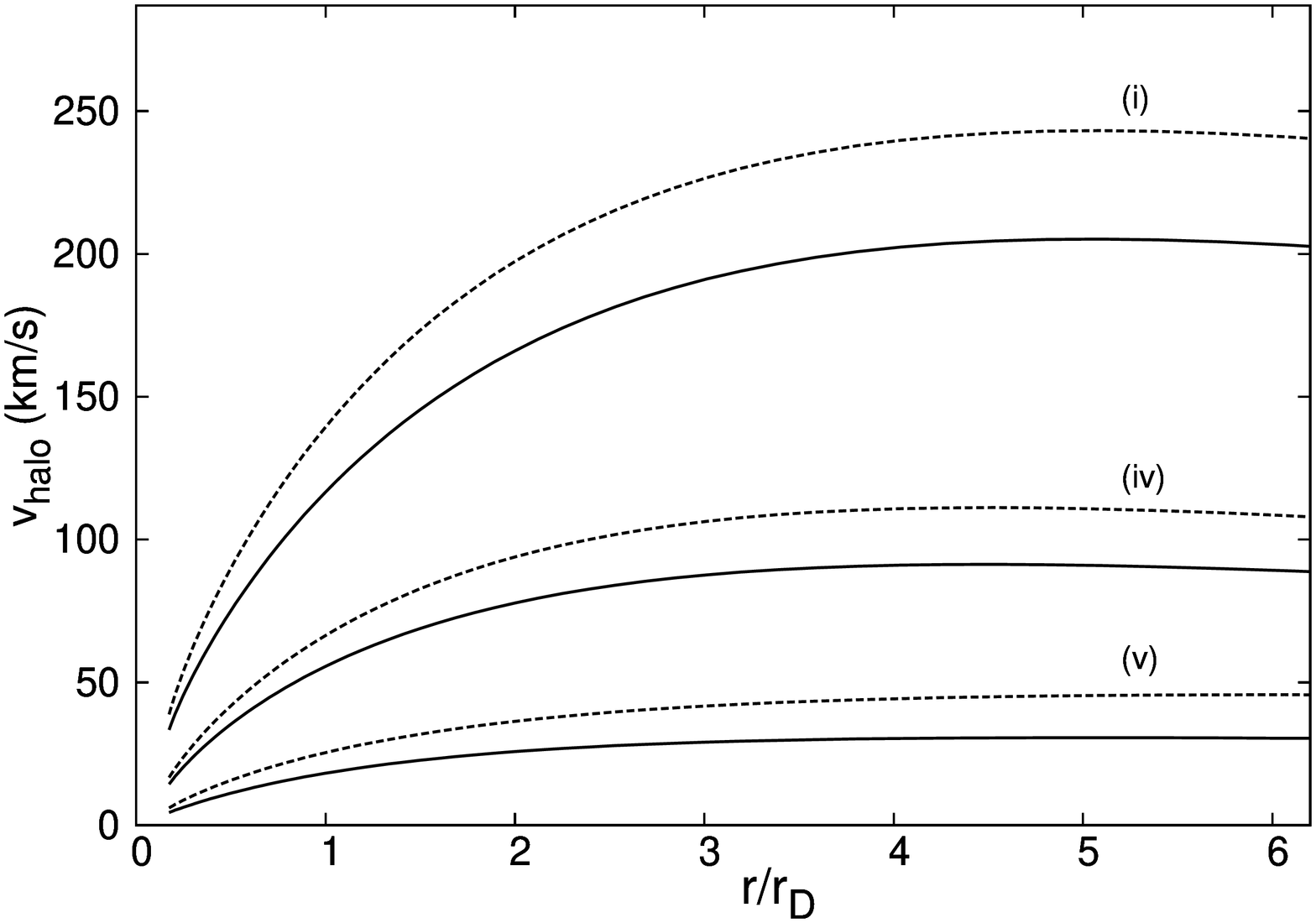}
    (b)
    \vspace{4ex}
  \end{minipage}
\vskip -1.0cm
\caption{
\small 
(a) Steady state rotation curves (halo+baryons) for the canonical baryonic parameters (i), (iv), (v) (solid line),
and a $r_D \to 2r_D$ variation (dashed line).
(b) The corresponding halo rotation curves.
}
\end{figure}

Consider now the effect of varying the  luminosity.
Specifically, we have computed the steady state solutions for
$M_{\rm FUV} \to M_{\rm FUV} \pm 0.8$,
with the other baryonic parameters ($m_{\rm baryon}$, $f_s$, $r_D$) 
unchanged from their canonical (Table 1) values.
This set of baryonic parameters, together with the canonical parameter choice, and
those with $r_D \to 2r_D$, provide a total of 20 distinct galaxy parameters.
While changing the FUV luminosity will strongly influence the normalization
of the halo rotational velocity, let us first look at the effect
on the shape of the rotation curve.

Figure 11 shows the
normalized halo rotational velocity, 
$v_{\rm halo}(r)/v_{\rm halo}(r_{\rm opt})$, $r_{\rm opt} = 3.2r_D$, for all 20 modelled galaxies.
Evidently,
the halo rotation curves 
that follow from the numerical solution of the steady state conditions, Eq.(\ref{SSX}),
have a near universal profile. This profile is reasonably consistent with the rotation
curve of Figure 2, which results from the simple analytic 
estimate of Eq.(\ref{r1x}).
As discussed in that context, the shape of the halo rotation curve reflects
the geometry
of the heating sources (SN distribution).
Also note that the universal profile for the normalized 
halo rotational velocity is not far from the
rotation curve measurements of actual galaxies, e.g. \cite{littlethings,sal17}.

\begin{figure}[t]
  \begin{minipage}[b]{0.5\linewidth}
    \centering
    \includegraphics[width=0.7\linewidth,angle=270]{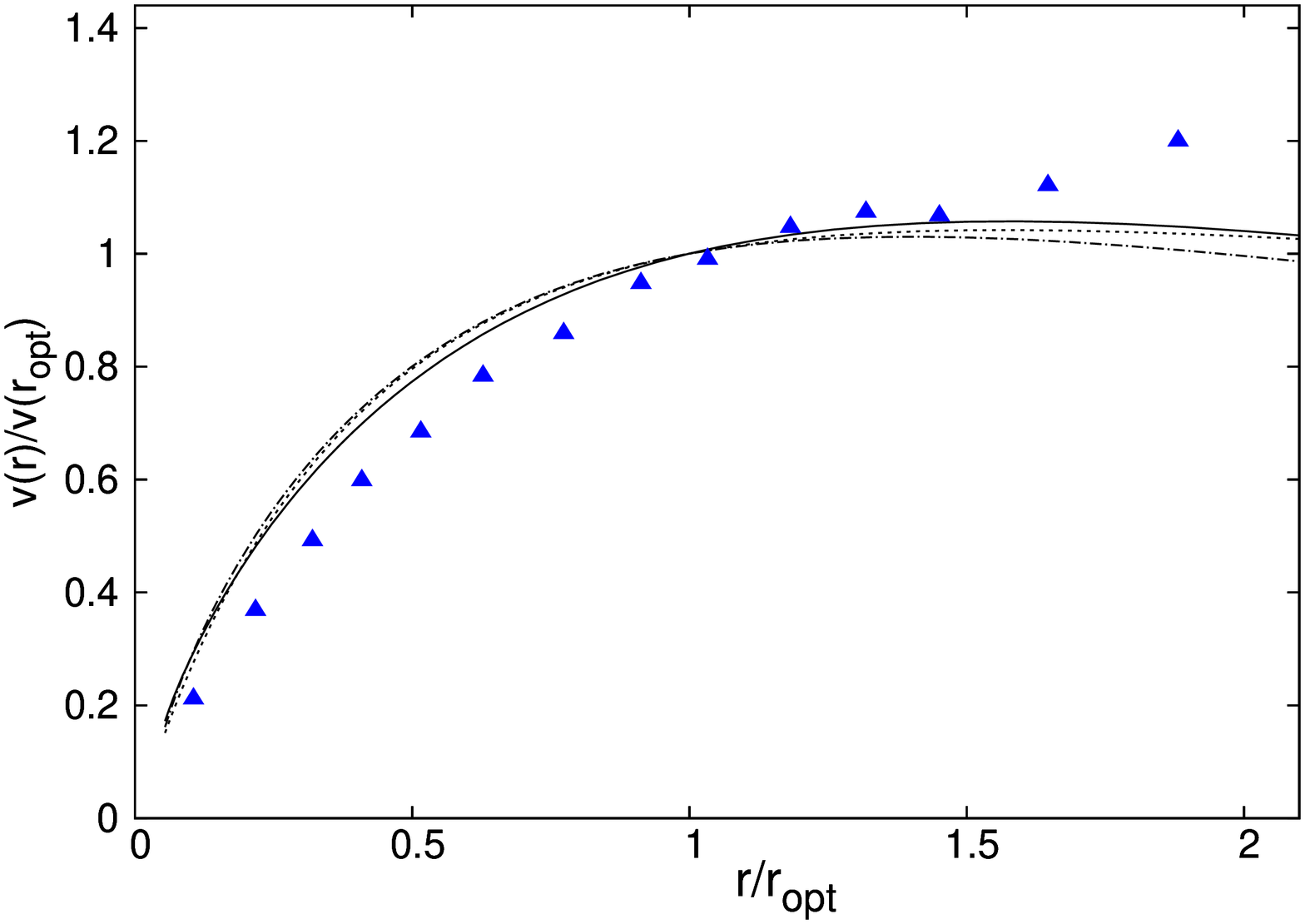}
     (a)
    \vspace{4ex}
  \end{minipage}
  \begin{minipage}[b]{0.5\linewidth}
    \centering
    \includegraphics[width=0.7\linewidth,angle=270]{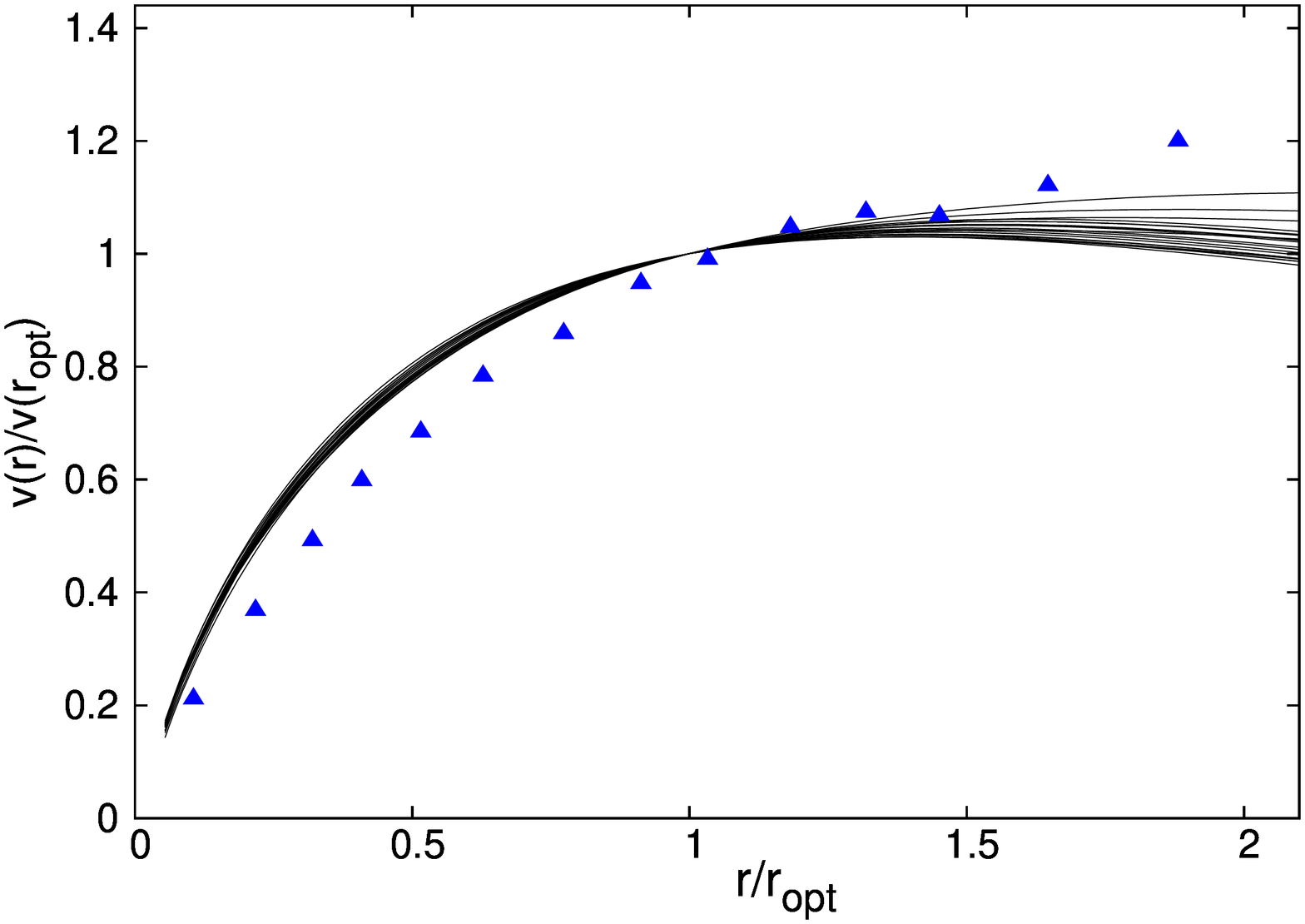}
    (b)
    \vspace{4ex}
  \end{minipage}
\vskip -1.0cm
\caption{
\small 
(a) Normalized halo rotational velocity, $v_{\rm halo}(r)/v_{\rm halo}(r_{\rm opt})$, 
calculated from the steady state solutions
for the canonical baryonic parameters (i) (solid line), (iv) (dashed-dotted line) and 
(v) (dotted line).
(b) The corresponding results
for all 20 modelled galaxies.
Triangles are the synthetic rotation curve derived from dwarf galaxies \cite{sal17}.
}
\end{figure}

\subsection{Normalization of the rotational velocity}

Consider now the
scaling of the normalization of the halo velocity.
Analytic considerations, leading to Eq.(\ref{ss1}),
suggest that the cooling function
can be related to $\kappa, \ r_D$ and $v_{\rm halo}^{\rm max}$:
\begin{eqnarray}
\Lambda \propto \frac{\kappa r_D}{[v_{\rm halo}^{\rm max}]^4} 
\ .
\end{eqnarray}
In the low density optically thin limit, the cooling function depends only on the
halo temperature. The temperature
depends sensitively on
the asymptotic rotational velocity and relatively weakly on 
the other parameters. Thus, simple analytic arguments
suggest that the combination $\kappa r_D/[v_{\rm halo}^{\rm max}]^4$  might 
be, at least approximately, a function of  $v_{\rm rot}^{\rm asym}$ only.
In any case, this assertion can easily be  checked numerically.

To make contact with observable quantities, we again make use of the expected
$\kappa \propto L_{\rm FUV} \propto 10^{-0.4 M_{\rm FUV}}$ scaling, which leads us to  
introduce the related quantity: 
\begin{eqnarray}
\stackrel{\sim}{\Lambda} \ \equiv \ \frac{r_D \ 10^{-0.4M_{\rm FUV}}  }{[v^{\rm max}_{\rm halo}]^4}
\label{10x}
\ .
\end{eqnarray}
For each set of galaxy parameters considered, $\stackrel{\sim}{\Lambda}$ 
can be determined
from the computed steady state
solutions.
The result of 
this exercise is shown in Figure 12, where we plot the  obtained $\stackrel{\sim}{\Lambda}$ values
versus the asymptotic rotational velocity, $v_{\rm rot}^{\rm asym}$.

\begin{figure}[t]
    \centering
\includegraphics[width=0.48\linewidth,angle=270]{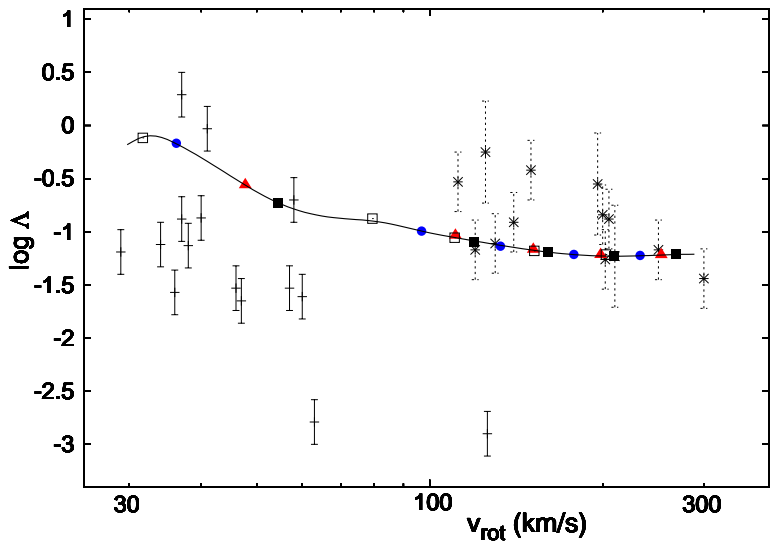}
\caption{
\small
$\log \stackrel{\sim}{\Lambda} \equiv -4\log[v_{\rm halo}^{\rm max} {\rm (km/s)}] +  
\log [r_D {\rm (kpc)}] - 0.4M_{\rm FUV}$
versus the asymptotic rotational velocity 
computed from the steady state solutions found.
Circles denote the baryonic parameters
of Table 1, triangles a $r_D \to 2r_D$ variation, and filled (unfilled) squares for a
$M_{\rm FUV} \to M_{\rm FUV} - 0.8$  ($M_{\rm FUV} \to M_{\rm FUV} + 0.8$) variation.
The solid line is an extrapolation of the computed solutions.
Also shown are the
$\log \stackrel{\sim}{\Lambda}$  values from THINGS spirals  \cite{things} 
and LITTLE THINGS dwarfs \cite{littlethings}.
}
\end{figure}

Also shown in the figure
are the values of $\stackrel{\sim}{\Lambda}$  for THINGS spirals \cite{things} and 
the classically shaped LITTLE THINGS dwarfs \cite{littlethings}.
The galaxy FUV absolute magnitudes were obtained from the
updated nearby galaxy catalogue \cite{table}, and 
corrected for internal and foreground extinction following \cite{uv3,uv2}.
The $r_D$ values for the dwarfs were obtained by fitting the 
FUV surface brightness profile (from \cite{Herr}) to a single 
exponential, while the $r_D$ values for spirals were extracted from 
a fit to their stellar rotation curves.
The maximum of the halo rotational velocity
was estimated from existing fits of rotation curves that used the 
isothermal profile \cite{littlethings,things}.
Table 3 summarizes the relevant quantities for the galaxies considered.

Indicative errors on the $\stackrel{\sim}{\Lambda}$ values for these galaxies
were estimated as follows.
For  the dwarfs  we took indicative  errors in 
$r_D$ and $v_{\rm halo}^{\rm max}$ of 10\%.
For the spirals, baryons typically contribute much more to the rotation velocity 
which increases the uncertainty in extracting $v_{\rm halo}^{\rm max}$.
We took indicative errors of 
$20\%$ for $r_D$, and
15\% for $v_{\rm halo}^{\rm max}$, except for the three spirals with the largest baryonic contribution
(NGC4736, NGC3031, NGC3521) for which this uncertainty was
taken to be 30\%.
Determining the absolute magnitude requires the distance to the galaxy 
(we took the distance values given in
\cite{littlethings} and \cite{things}), and we assumed a distance uncertainty 
for both dwarfs and spirals of 10\%.

\subsection{Deviant dwarfs}

In deriving the steady state solutions we assumed the scaling of halo heating
given by Eq.(\ref{sc99}), with $\kappa_{\rm MW}$ fixed so that $v_{\rm rot}^{\rm asym}$
roughly matched expectations for a Milky Way scale galaxy.
With this scaling and normalization, Figure 12 indicates that
the $\stackrel{\sim}{\Lambda}$ values of the dwarfs are offset from the 
theoretical curve;
most of the dwarfs considered have a $\stackrel{\sim}{\Lambda}$ value below the
value calculated from the steady state solutions.
The data is possibly consistent with the $\stackrel{\sim}{\Lambda} \propto \sqrt{\langle T_{\rm halo} \rangle} \propto v_{\rm rot}^{\rm asym}$
scaling, which is expected within a dissipative model where bremsstralung dominates the cooling
over all galaxy scales of interest. Within the framework of generic two component dark matter
models, along the lines of e.g. \cite{sunny1}, a large range of parameter space is  
anticipated, although with $\partial \Lambda/\partial T > 0$, the stability of the
steady state solution becomes a pressing issue (as will be discussed shortly).
Although it is tempting to look into this in more detail,
we shall postpone this for another time, and here continue within
the theoretically constrained mirror dark matter framework.

The offset of the dwarfs from the steady state $\stackrel{\sim}{\Lambda}$ solution curve
is not insignificant, but there is some `wiggle room'  arising 
from the various assumptions and approximations made.
The  list of things to consider would include: The use of the idealized 
spherically symmetric modelling,
simplistic modelling of the SN distribution (a single exponential  
is known to be a poor representation for many of the dwarfs, e.g. \cite{Herr}),
possible role of dark magnetic fields.
\footnote{
It might also be possible that SN sourced dark photons contribute to the ${\rm He'}$ ionization, a
situation that would require a substantial flux near the helium ionization energy. 
This could significantly reduce
the cooling losses due to line emission and thereby flatten out the cooling function. 
}
For the purposes of this paper, however, we shall consider only one possibility in detail,
namely, that the rate of halo heating for the dwarfs is significantly larger than anticipated from the 
FUV luminosity scaling. 
For example, $\kappa_{\rm MW}$ would be larger if the halo
were modelled with
a significant mirror metal component ($\gtrsim 0.1$ \% by mass).
It could also be possible that
the types of SN's in dwarfs, and their environment, 
might lead to a larger fraction of SN energy deposited into the halo.
To illustrate this scenario, we modelled a set of dwarf galaxies with 
the halo heating raised by a universal factor  of $\log\kappa \to \log\kappa + 0.8$. 
We calculated steady state solutions for several dozen distinct examples,
covering a wide range of baryonic parameters,
including: $-16 \le M_{\rm FUV} \le -7$, $0.05 \le r_D/{\rm keV} \le 1.2$.
Figure 13 shows the resulting $\stackrel{\sim}{\Lambda}$ values evaluated from 
these steady state solutions with
raised $\kappa$ values.

\begin{figure}[t]
    \centering
\includegraphics[width=0.48\linewidth,angle=270]{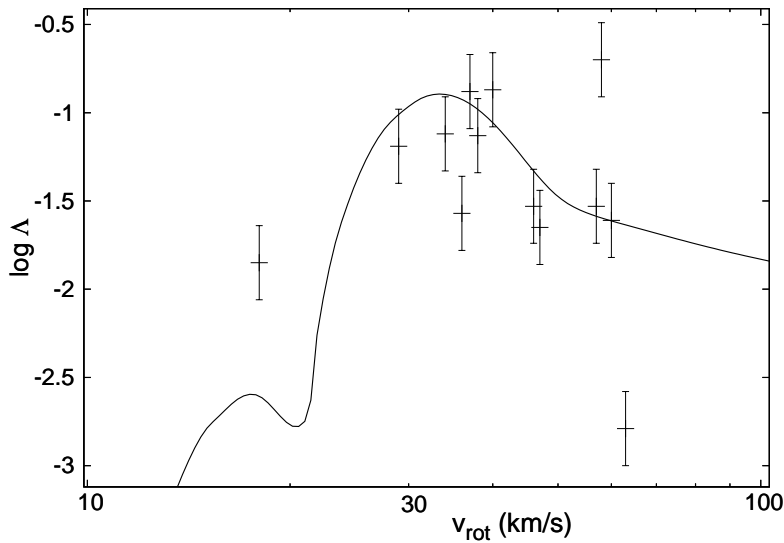}
\caption{
\small
$\log \stackrel{\sim}{\Lambda}$ values computed from the
steady state solutions 
with $\kappa$ universally increased [$\log\kappa \to \log\kappa + 0.8$]. 
Shown is the extrapolation of the computed solutions, together
with LITTLE THINGS dwarfs.
}
\end{figure}

The $\stackrel{\sim}{\Lambda}$ values for the dwarfs 
show some diversity. Roughly a third of the LITTLE THINGS dwarfs
have not been considered here. Their rotation curves are not classically
shaped. They might have dissipative halos undergoing perturbations
of some kind, some may have collapsed halos; 
their interpretation is unclear.
Of the dwarfs considered, 
there are three galaxies with relatively high values
of $\stackrel{\sim}{\Lambda}$ (DDO50, NGC1569, NGC2366) 
and two with relatively low values (DDO101,  NGC3738).
The three dwarfs with the highest $\stackrel{\sim}{\Lambda}$
values are known to be starburst galaxies, which feature 
oscillating star formation rates (SFR) with a period of order 100 Myr \cite{starburst}.
A dissipative halo around such galaxies would not be expected 
to be in a steady state configuration, but expanding and contracting
in phase with the SFR, and maybe even driving the oscillating SFR. 
In fact the observations indicate \cite{starburst} 
that the SFR for these three dwarfs is
near the peak of this cycle, suggesting that the halo would have
${\cal H} > {\cal C}$. This  is consistent with their higher $\stackrel{\sim}{\Lambda}$ values.
The  two dwarfs with very low $\stackrel{\sim}{\Lambda}$ values could be 
interpreted as dwarfs with a halo that is no longer dissipative.
That is, the halo is not
in the form of a plasma, but in a collapsed configuration, e.g. in a disk containing dark stars.

\begin{figure}[t]
  \begin{minipage}[b]{0.5\linewidth}
    \centering
    \includegraphics[width=0.7\linewidth,angle=270]{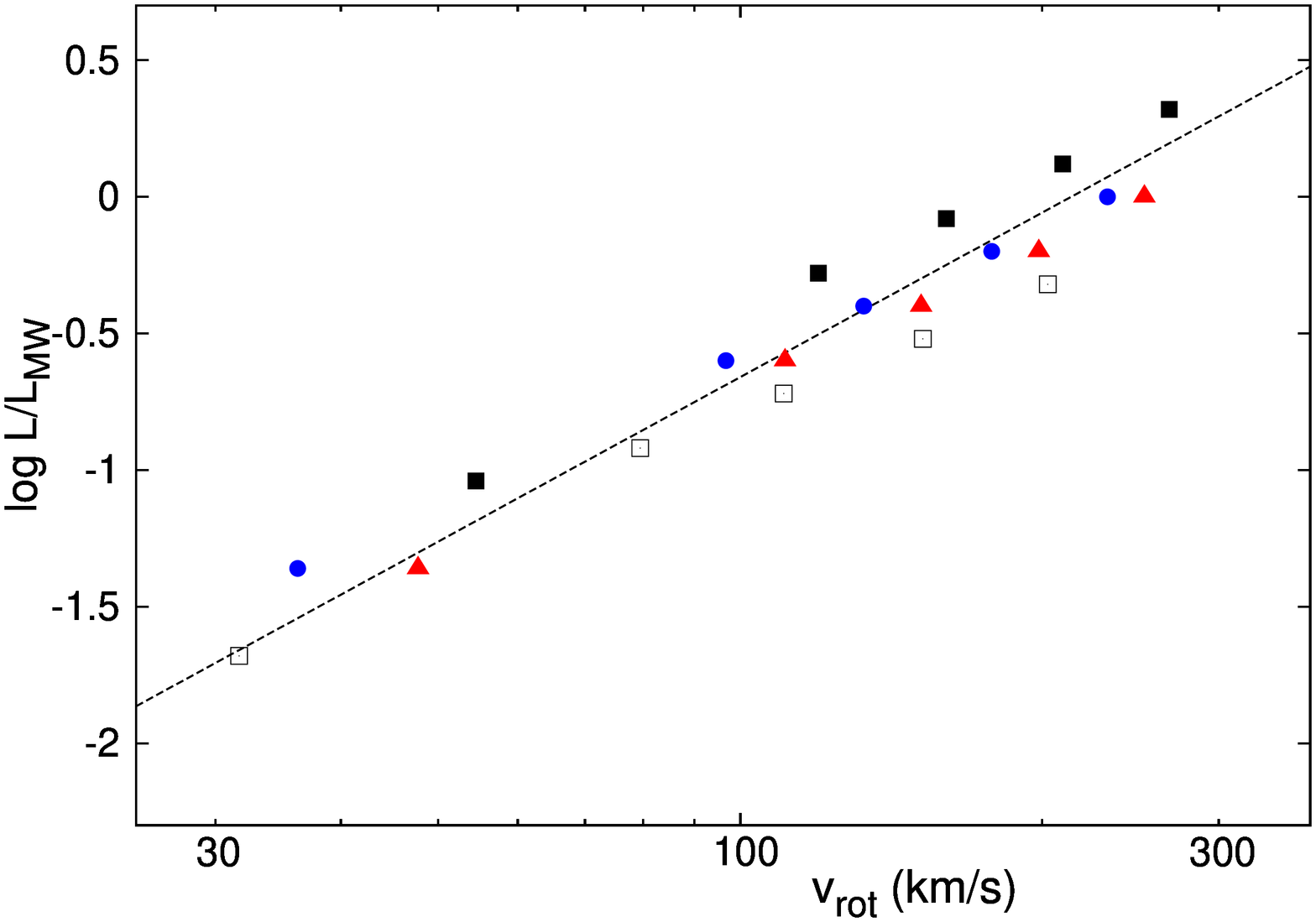}
     (a)
    \vspace{4ex}
  \end{minipage}
  \begin{minipage}[b]{0.5\linewidth}
    \centering
    \includegraphics[width=0.7\linewidth,angle=270]{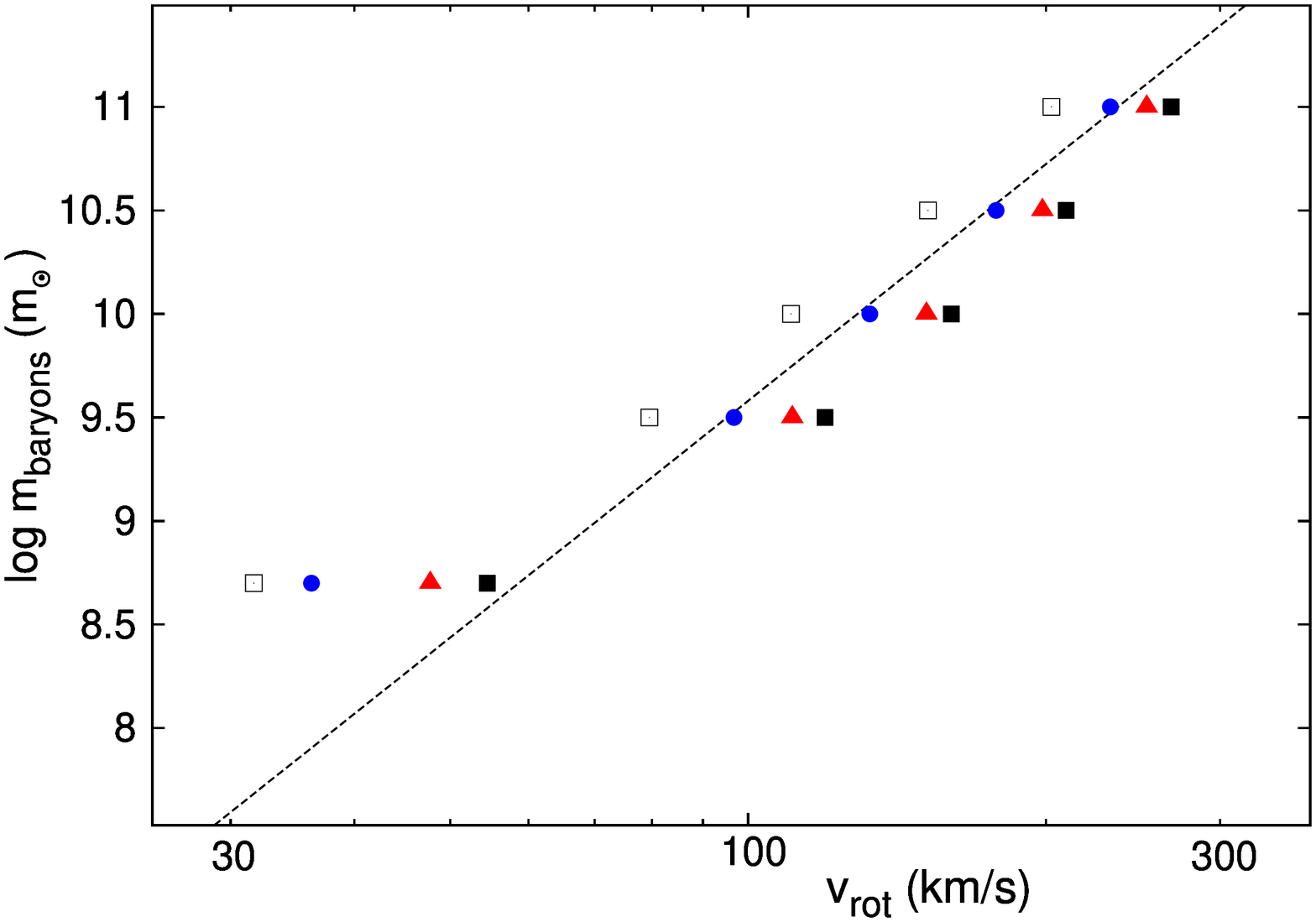}
    (b)
    \vspace{4ex}
  \end{minipage}
\vskip -1.0cm
\caption{
\small
(a)  $L_{\rm FUV}/L^{\rm MW}_{\rm FUV}$ and (b) $m_{\rm baryon}$ 
versus the asymptotic rotational velocity 
computed from the steady state solutions found. 
Circles  are the results for the canonical galaxy set of
Table 1, triangles for a $r_D \to 2r_D$ parameter variation, and filled (unfilled) squares  for a
$M_{\rm FUV} \to M_{\rm FUV} - 0.8$ ($M_{\rm FUV} \to M_{\rm FUV} + 0.8$)  variation.
The dashed lines are the power laws:  $L_{\rm FUV} \propto [v_{\rm rot}^{\rm asym}]^{2.0}$ and
$m_{\rm baryon} \propto [v_{\rm rot}^{\rm asym}]^{3.8}$.
}
\end{figure}

There is one very dim dwarf galaxy, DDO216, with low asymptotic
rotational velocity, $v_{\rm rot}^{\rm asym} \lesssim 20$ km/s \footnote{
There was one other very small dwarf in the LITTLE THINGS sample,
DDO210. Unfortunately,
the rotation curve of that dwarf was too poorly constrained
to be useful.
},
so low in fact, that the mean temperature in the radial region of interest, $r \lesssim 5r_D$,
would be {\it below} the temperature of the ${\rm He \ II}$ line emission peak.
This peak is shown in Figure 1a for an optically thin halo, and is also
evident in Figure 13, at $v_{\rm rot}^{\rm asym} \sim 30$ km/s.
\footnote{The cooling function, $\Lambda$, as defined for Figure 1,
and the quantity $\stackrel{\sim}{\Lambda}$ defined in Eq.(\ref{10x})
are closely connected with each other.
The latter can be considered as a kind of average over the former,
with some corrections. E.g. the cooling function $\Lambda$ was calculated
in the optically thin limit, while  the quantity $\stackrel{\sim}{\Lambda}$
includes corrections to the ionization state from halo reabsorption of
radiation.
}
As will be discussed shortly, 
the steady state solution could be unstable in the 
temperature region where $\partial \Lambda/\partial T$ is large in 
magnitude and positive in sign.
However, the asymptotic rotational velocity for DDO216 is so low, that 
the $\stackrel{\sim}{\Lambda}$ value for this
dwarf could conceivably be in the `plateau' region
below the ${\rm He \ II}$  line emission peak. If so,
DDO216 might be a rare example of a dwarf
with a dissipative halo
with $v_{\rm rot}^{\rm asym}$ below the ${\rm He \ II}$ peak.



In this picture, the rotation curve normalization 
is set by equating the heating and cooling rates, and 
this leads to the $\stackrel{\sim}{\Lambda}$ curves
of Figures 12, 13.
This would presumably be the `fundamental' relation underpinning the empirical Tully Fisher relations \cite{tf,btf,btfnew}.
The latter, could be interpreted as approximate relations.
In Figure 14a [14b] we plot $L_{\rm FUV}$ [$m_{\rm baryon}$] versus $v_{\rm rot}^{\rm asym}$ for
the modelled galaxies.
The results for the 20 modelled galaxies fall in a band, instead of a line, as the 
`fundamental' relation is the one involving $\stackrel{\sim}{\Lambda}$ rather than $L_{\rm FUV}$ or $M_{\rm baryon}$.

The results shown in Figure 14 were computed from the steady state solutions derived with
the original $\kappa$ scaling of Eq.(\ref{sc99}), $\kappa_{\rm MW} = 2\times 10^{44}$ erg/s, for
both the spirals and the dwarfs (i.e. these figures do not include the ${\rm log} \kappa \to {\rm log} \kappa +  0.8$
adjustment for the dwarfs).
With this original $\kappa$ scaling, 
Figure 14b indicates that the numerical solution for the dwarfs are somewhat offset from the linear extrapolation of the numerical
solution for the spirals, when baryon mass is considered. 
On the other hand, Figure 14a indicates that the that the numerical solution for the dwarfs is in agreement with the linear extrapolation of the 
numerical solution for the spirals, when luminosity is instead examined.
This is in contrast to 
observations which are consistent with a linear baryonic mass relation (Baryonic Tully Fisher relation) \cite{btf,btfnew}, 
and is another manifestation of the slight offset of the normalization of the rotational velocity of the dwarfs identified in Figure 12, 
and discussed earlier.
If the $\kappa$ is adjusted for the dwarfs, ${\rm log} \kappa \to {\rm log} \kappa +  0.8$,
then the asymptotic rotational velocity from the numerical solution increases by around a factor of two,
and the deviant behaviour is corrected.

\section{Stability considerations}

We did not need to know anything about the galaxy's history to determine the steady state solutions.
Naturally, one would love to know all about the dynamical evolution 
of the halo, i.e. how the halo got to the steady state configuration.
One would also like to check that the steady state configuration is stable.
Perturbations about the steady state solution should not lead to runaway expansion or contraction.
These are all important issues that have yet to be addressed.
To fully address these issues would require the solution of the full time-dependent Euler 
fluid equations
and also to model the baryonic SFR in the evolving environment.
No such study will be undertaken here, instead a  qualitative description 
will be attempted.

Dissipative dark matter halos are coupled to the baryons via SN sourced heating. 
The baryons are also coupled to the halo via gravity. 
One would expect that as the halo expands, the ordinary matter will also
expand in the weakened gravity.
This will have a strong effect on the SFR, which is known 
to sensitively depend on the average baryon gas density \cite{ken,Schmidt}.
This reasoning suggests that an
expanding halo will have a reducing SFR, while a contracting one will have increasing
SFR. Evidently, the implications of expanding and contracting halos are quite nontrivial.
Nevertheless, a plausible story for how galaxies 
may have evolved emerges.

Once upon a time,
prior to the onset of significant star formation,
the halo may have been in a more collapsed configuration compared with 
its present expanded distribution.
This is the expected consequence of dissipation without any compensating heat source.
At that time, the baryons would also be much more compressed than they are now, as
the condensed halo would dramatically increase the gravity.
As baryons collapse and cool
they can form stars, and in this dense  environment the star formation 
rate can potentially be extremely 
high. At some point, supernovae would occur and the heating of the halo would begin.
For the purposes of this qualitative discussion, consider the
heating and cooling rates suitably averaged over the halo,
$\langle {\cal H} \rangle$, $\langle {\cal C} \rangle$.
In this early phase, $\langle {\cal H } \rangle > \langle {\cal C } \rangle$, and the
halo rapidly expands. This in turn weakens the gravity and decreases the SFR.
Note that the
timescale over which large stars evolve to form SN
is relatively short ($\sim 10$ Myr) when compared to the timescale of the halo evolution 
($\sim 100$ Myr or more).

A necessary condition for the system to be able to evolve to a steady state configuration
is that $\langle {\cal H } - {\cal C } \rangle$ reduces as the halo expands.
In fact $\langle {\cal H} - {\cal C}\rangle$ will need to be able to reduce until
$\langle H \rangle = \langle C \rangle$. If this does indeed happen,
then in this initial expansion phase 
the halo would presumably continue to expand past this near equilibrium configuration, due
to the momentum of the fluid.
Just after it overshoots the equilibrium configuration,
$\langle {\cal H } - {\cal C } \rangle < 0$, and the expansion of the halo will be slowed.
The halo density will oscillate around the equilibrium configuration.
Such oscillations would presumably be damped, and the system
would eventually relax to the steady state configuration. In the steady state,
not only does $\langle {\cal H} \rangle = \langle {\cal C } \rangle$, 
but this balancing must happen {\it locally}
so that ${\cal H} = {\cal C}$.
The observable effects of the evolving halo could be seen on the SFR as this tracks the gravity.
In fact,
there is some  evidence supporting this picture from studies of 
the SFR history of the Milky Way \cite{snaith}, including
an initial period with rapid SFR, followed by a dip in SFR 
(the expected distinctive feature of a halo that
has expanded past the equilibrium configuration), and subsequently the SFR relaxes to
a near constant.

Naturally, it would be useful to check that
$\langle {\cal H } - {\cal C}\rangle < 0$ does indeed arise if the halo were to expand past the 
$\langle {\cal H} \rangle = \langle {\cal C} \rangle$ 
configuration.
This stability condition is not so easy to verify. There  are three components which
we now describe. 
Firstly, the
SFR will decrease as the halo expands, due to the effect of the weakening gravity, as discussed above.
This will reduce $\langle {\cal H } \rangle$.
Secondly,
as the halo expands the density reduces which reduces also  $\langle {\cal C } \rangle$ as 
${ \cal C} \propto \rho^2$.
Thirdly, this effect can be compensated, at least in part, by an increase in the cooling function, $\Lambda$.
As the halo expands it cools. If the expansion is slow enough then the halo will remain in hydrostatic equilibrium,
but since $v_{\rm rot}$ is reduced in the weakened gravity, 
the temperature will be lower as $T_{\rm halo} \propto [v_{\rm rot}^{\rm asym}]^2$.
If the halo is in a temperature region where
$\partial \Lambda/\partial T < 0$, then this will lead to an increase in $\langle {\cal C}\rangle$. 
From Figure 4b, we see that the temperature range of interest
is roughly: $\langle T_{\rm halo} \rangle  \approx 0.3  - 0.007$ keV,
and the cooling function, as shown in Figure 1 for the optically thin limiting case, does indeed feature a 
negative slope: $\partial \Lambda/\partial T < 0$ in this
temperature range. 
Such stability considerations
might ultimately be able to provide an explanation for the observed physical scale of star forming galaxies:
$v_{\rm rot}^{\rm asym} \sim 30$ km/s for the smallest dwarfs to around $300$ km/s for the largest
spirals.
[In addition, a region of stability might possibly exist in the plateau below the  He II  line emission peak
where $v_{\rm rot}^{\rm asym} \sim 20$ km/s.]
This is all very interesting, but
also rather qualitative, and so it is still unclear
whether $\langle {\cal H } - {\cal C } \rangle$ does indeed reduce when the halo 
expands past its equilibrium
configuration. Much more work is needed to explore this and other dynamical issues.

\section{Discussion}

Within the mirror dark matter model, and dissipative dark matter models in general, halos 
around galaxies with active star formation (including spirals and gas rich dwarfs)
are dynamical: they
expand and contract in response to 
heating and cooling processes. 
This dynamics allows halo properties to be strongly influenced by baryons as ordinary
Type II supernovae can provide the dominant heat source.
This seems to provide a rather simple explanation for the baryon - dark matter connection
that has been identified from rotation curve measurements, 
and discussed extensively in the literature, e.g.\cite{salucci,MC,DS,Lelli,stacy,sal17}.

The kinetic mixing interaction provides the bridge between the ordinary particles and their mirror sector partners.
This interaction, even if very tiny ($\epsilon \sim 10^{-9}-10^{-10}$),
transforms Type II SN into powerful dark sector heat sources. 
In the SN core, this interaction generates an expanding  energetic plasma 
of dark matter particles and dark radiation,
with total energy of around
$\sim 10^{53}$ erg for the kinetic mixing strength considered.
The dark plasma evolves into a relativistic fireball and
sweeps up the surrounding halo mirror baryons as it expands.
The energy of the fireball is expected to be transferred  efficiently into the mirror baryons.
Ultimately, this flow is decelerated, and part of this kinetic energy
is converted into thermal energy which can radiatively cool producing dark radiation
(analogous to the fireball model of GRB).
The end result is that 
the SN sourced energy is transmitted to the halo in two distinct ways: via dark photons, and also 
via the heating of mirror baryons in the SN vicinity.

Dissipative dark matter halos can be modelled as a fluid governed by Euler's equations.
Around sufficiently isolated and unperturbed galaxies 
the halo can relax to a steady state configuration, where 
heating and cooling rates locally balance and hydrostatic equilibrium prevails.
These steady state conditions can be solved to obtain 
the current physical properties, including the halo density and temperature
profiles, for model galaxies.
We have considered idealized spherically symmetric galaxies within the mirror dark particle model 
as in the earlier paper, Paper I, 
but we have assumed that the local halo heating in the SN vicinity dominates over radiative sources.
With this assumption, physically interesting steady state solutions arise which we compute
for a representative range of model galaxies. 
These solutions, it turns out, 
have physical properties that closely resemble the empirical properties of disk galaxies.

Naturally, one can take issue with the idealized spherical symmetry approximation.
Certainly it would be important to obtain steady state solutions without resorting
to such a simplification. To this end, one could extend the methodology 
developed here, and in Paper I, to an azimuthally symmetric disk geometry.
However, there is a slight complication if local halo heating in the SN vicinity dominates
over radiative sources:
the SN sourced heating would presumably
require conduction, convection, and collective halo motion to transfer this heat
to the bulk of the halo.
These processes would need to be modelled if the steady state solutions are to be at all
realistic for the disk geometry case.
This can be contrasted with the 
idealized spherical symmetric systems that have been considered here, in which the 
SN sourced heating is spread throughout
the whole volume, and an almost isothermal halo results in the region of 
most interest, $r \lesssim 5r_D$. Due to this fortunate circumstance, 
consideration of such
heat transfer processes is
not essential; and in any case, they were not included in the analysis presented in this paper.

The halo heating mechanism discussed provides physically interesting solutions,
and it also greatly simplifies the emerging picture.
In particular no significant mirror metal component is required in the halo.
The halo might consist predominately of only three components ${\rm H', \ He', \ e'}$, with 
trace amounts of heavier elements.
If this is the case there will be important implications for direct detection experiments.
Interactions of the light ion components (${\rm H', \ He'}$)
with heavy nuclei (Xe) should typically produce sub-keV energy deposits in such detectors, 
likely below current detector energy thresholds.
Constraints considered in e.g.\cite{mypaper}, from the 
sensitive XENON experiments, e.g. \cite{xenon,panda}, 
can thereby be ameliorated. 
Such experiments might be more sensitive to mirror 
electron scattering off target electrons.
However, the event rate is quite uncertain for several reasons.
First, 
the rate depends sensitively on the halo dark matter density and temperature at the
detector's location.
The corrections to these quantities that will inevitably arise 
when the SN distribution is modelled with disk geometry
could be significant, especially as the detector is located 
in the galactic disk.  For instance, the density may 
be much lower, and temperature much higher than that found in the spherically symmetric
steady state solutions discussed here. Second,
shielding of the detector from the halo wind by mirror dark matter captured within the Earth
can potentially suppress dark matter signals. It can also enhance the annual
modulation amplitude, produce large diurnal variations, etc \cite{plasmadm,f98}. There is some potential to
explain the long standing DAMA annual modulation signal \cite{dama1,dama}, although at the present
time it is unclear if this
remains viable given the constraints from other  experiments.

\vskip 2.1cm
\noindent
{\large \bf Acknowledgments}

\vskip 0.3cm
\noindent
This work was supported by the Australian Research Council.
The author would like to thank the referee for his/her comments which lead to improvements to both
this paper and Paper I.


\begin{table}[b]
\begin{center}
\begin{tabular}{c c c c c c c}
\hline\hline
Galaxy & $m_{\rm baryon}$   & $r_D$  & $M_{\rm FUV}$  & $v_{\rm halo}^{\rm max}$ & $v_{\rm rot}^{\rm asym}$ & $\log\stackrel{\sim}{\Lambda}$ \\
       &   $[10^8 m_\odot]$ & (kpc) &  (mag)        &  (km/s) & (km/s)   &  
\\
\hline
\rule{0pt}{4ex}
DDO43 & 2.3 & 0.55 & -13.14  & 34 & 38 & -1.13      \\
DDO50 & 14.0 & 1.07 & -15.42  & 30 & 37 & 0.29     \\
DDO52 & 4.1 & 1.17 & -13.36  & 57 & 60 & -1.61      \\
DDO53 & 0.80 & 0.34 & -12.51  & 27 & 29 & -1.19      \\
DDO87 & 3.5 & 1.45 & -13.09  & 54 & 57 & -1.53      \\
DDO101 & 0.93 & 0.56 & -11.59  & 62 & 63 & -2.79      \\
DDO126 & 1.9 & 0.77 & -13.39  & 34 & 37 & -0.88      \\
DDO133 & 1.5 & 0.63 & -13.00  & 43 & 46 & -1.53      \\
DDO154 & 3.7 & 0.58 & -13.10  & 46 & 47 & -1.65     \\
DDO216 & 0.21 & 0.19 & -9.48 & 17 & 18 & -1.85         \\
NGC1569 & 4.1 & 0.33 & -16.80  & 37 & 41 & -0.03      \\
NGC2366 & 11.9 & 1.09 & -15.32  & 52 & 58 & -0.70      \\
NGC3738 & 2.5 & 0.46 & -14.72  & 130 & 126 & -2.90      \\
WLM & 0.92 & 0.40 & -12.51  & 35 & 36 & -1.57      \\
Haro29$^*$ 
 & 0.94  & 0.34 & -13.68  & 34 & 34 & -1.12      \\
Haro36 & 1.7 & 0.40 & -14.61  & 38 & 40 & -0.87      \\
\hline
\rule{0pt}{4ex}
NGC925 & 130 & 2.1 & -18.28   & 110 & 112 & -0.53       \\
NGC2403 & 64 & 1.6 & -17.85   & 130 & 130 & -1.11      \\
NGC2841 & 1700 & 4.4 & -18.77   & 250 & 300 & -1.44       \\
NGC2903 & 203 & 1.9 & -18.45   & 170 & 202 & -1.26      \\
NGC3031 & 1030 & 2.4 & -18.08   & 110 & 196 & -0.55      \\
NGC3198 & 310 & 2.8 & -18.97   & 130 & 150 & -0.42     \\
NGC3521 & 1500 & 2.8 & -18.37   & 180 & 210 & -1.23     \\
NGC3621 & 240 & 1.7 & -18.29   & 130 & 140 & -0.91      \\
NGC4736 & 280 & 1.3 & -16.86   & 60 & 125 & -0.25      \\
NGC5055 & 1600 & 3.2 & -18.57   & 160 & 205 & -0.88      \\
NGC6946 & 780 & 3.0 & -18.74   & 160 & 200 & -0.84      \\
NGC7331 & 2300 & 2.4 & -18.96   & 190 & 250 & -1.17     \\
NGC7793 & 34 & 0.9 & -17.18   & 100 & 120 & -1.17     \\
\hline\hline
\end{tabular}
\caption{\small                     
Some properties of the 
considered LITTLE THINGS dwarfs and  THINGS spirals. Baryon mass, 
Disk scale length, FUV absolute magnitude (corrected for extinction),
maximum halo rotation velocity, asymptotic rotation velocity, and 
$\log \stackrel{\sim}{\Lambda}
 \equiv -4\log[v_{\rm halo}^{\rm max} {\rm (km/s)}] +  \log [r_D {\rm (kpc)}] - 0.4M_{\rm FUV}$.
}
\vskip 2.3cm
\end{center}


\footnotesize{$^*$ For Haro29, the small peak at $r = 3.6$ kpc 
in Fig. A.75 of \cite{littlethings} was not included
in estimating $v_{\rm halo}^{\rm max}$, while for Haro36 the rising rotation curve
at the largest measured radii was also ignored.
}
\end{table}

\end{document}